\documentclass[aps, prb, notitlepage, citeautoscript, superscriptaddress, eqsecnum, reprint]{revtex4-2}

\usepackage{xr}
\usepackage{amsmath}
\usepackage{amsthm}
\usepackage{amsfonts, amssymb,amsxtra}
\usepackage[]{graphicx}
\usepackage{grffile}
\usepackage{amsfonts}
\usepackage{framed}
\usepackage{bbm}
\usepackage{braket}
\usepackage{xcolor}
\usepackage{mathtools}
\usepackage{LatexCommands}
\usepackage{booktabs}

\usepackage[colorlinks=true]{hyperref}
\hypersetup{
    unicode=false,          % non-Latin characters
    pdftoolbar=true,        % show Acrobat
    pdfmenubar=true,        % show Acrobat
    pdffitwindow=false,     % window fit to page when opened
    pdfstartview={FitH},    % fits the width of the page to the window
    pdftitle={},    % title
    pdfauthor={},     % author
    pdfsubject={},   % subject of the document
    pdfcreator={},   % creator of the document
    pdfproducer={}, % producer of the document
    pdfkeywords={} {} {}, % list of keywords
    pdfnewwindow=true,      % links in new window
    colorlinks=true,       % false: boxed links; true: colored links
    linkcolor=blue, %red,          % color of internal links (change box color with linkbordercolor)
    citecolor=blue,        % color of links to bibliography
    filecolor=blue,      % color of file links
    urlcolor=blue           % color of external links
}

\begin{document}

\title{
Efficient Magic State Factory Via Transversal Non-Clifford Gate}

\author{I-Chi Chen}
\email{ichen@lanl.gov}

~\affiliation{Los Alamos National Laboratory, Computing and Artificial Intelligence Division, Los Alamos, NM, USA}

\author{Hrushikesh Pramod Patil}
\email{hpatil2@ncsu.edu}
~\affiliation{Department of Electrical and Computer Engineering, North Carolina State University, Raleigh, NC, USA}

\author{Huiyang Zhou}
~\affiliation{Department of Electrical and Computer Engineering, North Carolina State University, Raleigh, NC, USA}

\author{Andrew Sornborger}
~\affiliation{Los Alamos National Laboratory, Computing and Artificial Intelligence Division, Los Alamos, NM, USA}
\begin{abstract}

\noindent
Magic-state preparation is a central component of fault-tolerant quantum computing. Recent theoretical and experimental successes in code-switch-based magic-state preparation have underscored the promise of these methods for quantum error correction. Similarly, magic-state cultivation has likewise been demonstrated in both numerical and experimental settings. However, a thorough comparison between magic-state cultivation and code-switch-based magic-state factories is still missing. In this work, we carry out end-to-end simulations of magic-state preparation using code switching and compare its resource requirements and performance against magic-state cultivation. As part of this analysis, we develop a lattice-surgery protocol for transfer between the doubled color code and the rotated surface code. We extend the complete code-switching protocol to the $d=5$ doubled color code and perform the corresponding end-to-end simulations. Finally, we propose two fault-tolerant magic-state preparation protocols that combine phase-kickback checks with a transversal non-Clifford gate.
\end{abstract}
\date{\today}

\maketitle

%\begin{itemize}
%    \item Point out the difference between T gate and S state cultivation (Showing the results with idling noise and maybe w/o)
%    \item Code switch might have better scaling for T gate and S gate (Showing ungrown results for d=2 and d=3 ) 
%    \item Show the implementation of d=2 ([[10,1,2]] code and [[7,1,3]] code), d=3 ([[15,1,3]] code and [[7,1,3]] code) d=5 ([[49,1,5]] code and [[17,1,5]] code) %([[53,1,5]] code and [[19,1,5]] code). I will try both grafting and lattice surgery.  (Grafting: need to switch to 2D color code. LS: can shift to a 2D surface %code directly for d=3, 3D color code works for d=3.)

%    \item New protocol: cultivation ($\ket{+}_L$ check) + Transversal T (+ switch) ?
%    Growing via lattice surgery and grafting. 
%\end{itemize}
\section{Introduction}
The realization of a universal fault-tolerant quantum computer (FTQC)~\cite{Shor1996,Campbell2017,Yoder2016}, which could provide exponential speed-ups for some algorithms relative to a classical computer, has become the focus of quantum information science. Quantum error correction (QEC) codes have been touted as the way to achieve FTQC.
An important component for QEC codes is a transversal gate set. Transversal gates act independently across corresponding physical qubits in logical code blocks to create a corresponding logical operation without error. Transversal gate sets greatly reduce the complexity of implementing fault-tolerant operations, which is pivotal to the practical implementation of fault-tolerant computations. 

Unfortunately, there is no free lunch when it comes to a fully transversal and universal fault-tolerant gate set. According to the Eastin-Knill theorem~\cite{Eastin2009}, no single quantum error code has a complete universal gate set that can be implemented transversally. Generally, for specific codes, Clifford gates can be implemented using lattice surgery (LS)~\cite{Horsman2012, Litinski2019} or transversal gates~\cite{Nielsen_Chuang_2010}. Non-Clifford gates, like the T gate, are required to complete a universal gate set but are generally not implemented transversally for the same codes. The most common way to implement a non-Clifford gate is to prepare a magic state (typically probabilistically) and teleport it into the arbitrary logical state as the logical operation. Alternatively methods, such as code switching (CS)~\cite{Bombin2016} between QEC codes, can also be used to complete a universal gate set. In this context, one switches between a code supporting all transversal Clifford gates and another code that supports a transversal non-Clifford gate. 
With CS techniques, one can achieve the universal fault-tolerant gate set. An example is CS between the $2$D color code and $3$D color code or double color code, which has transversal Clifford gates~\cite{Dehaene2003,Chamberland2020,Wu2023} and transversal $T$ gate (non-Clifford)~\cite{Bombin2007}, respectively. This approach to QEC has led to CS being studied extensively
%Hence, there is a lot of research on CS
between codes that support the aforementioned criteria~\cite{Kubica2015,Anderson2014,Bombin2015,Butt2024,Heussen2025,Pogorelov2025}. 

The fault-tolerant state preparation of a $3$D color code state is complicated. Although the $2$D color code has potentially better performance per physical qubit than its competitor, the rotated surface code, decoding on the $2$D color code has proven to be difficult. Only the maximum likelihood~\cite{Bravyi2014}, tesseract~\cite{beni2025}, and Vibe~\cite{Koutsioumpas2025} decoders, which have longer runtime, allow color codes to beat the surface code in terms of the performance per physical qubit. Their long runtime makes the practical implementation of $2$D color code infeasible. In contrast, surface codes with the minimum-weight perfect matching (MWPM) decoding can achieve good performance~\cite{Higgott2025} with much lower runtime overhead. However, the initialization and accurate decoding of the $3$D color code or doubled color code remains a challenge. This limits the potential for building a universal fault tolerant quantum computer solely based on CS between $2$D color code and $3$D color code.   

Instead of directly constructing a universal quantum computer with just the $2$D and $3$D color codes, Ref.~\cite{Daguerre2025} directly uses these two codes as a magic state factory, which allows one to discard non-trivial syndrome measurement outcomes~\cite{Daguerre25}. The corresponding protocol has also been successfully implemented on trapped-ion quantum hardware~\cite{Daguerre2025}. However, extending the protocol to $d=5$ is complicated because of its large qubit counts. In particular, fault-tolerantly preparing the $d=5$ tetrahedral color code requires $65$ qubits~\cite{Butt2025}. Alternatively, one can also use the $d=5$ doubled color code using double construction with the $2$D $d=5$ color code. Nevertheless, fault-tolerant initialization of the doubled color code remains a challenge. 

\begin{figure}[t]
\centering
\includegraphics[width=1\linewidth]{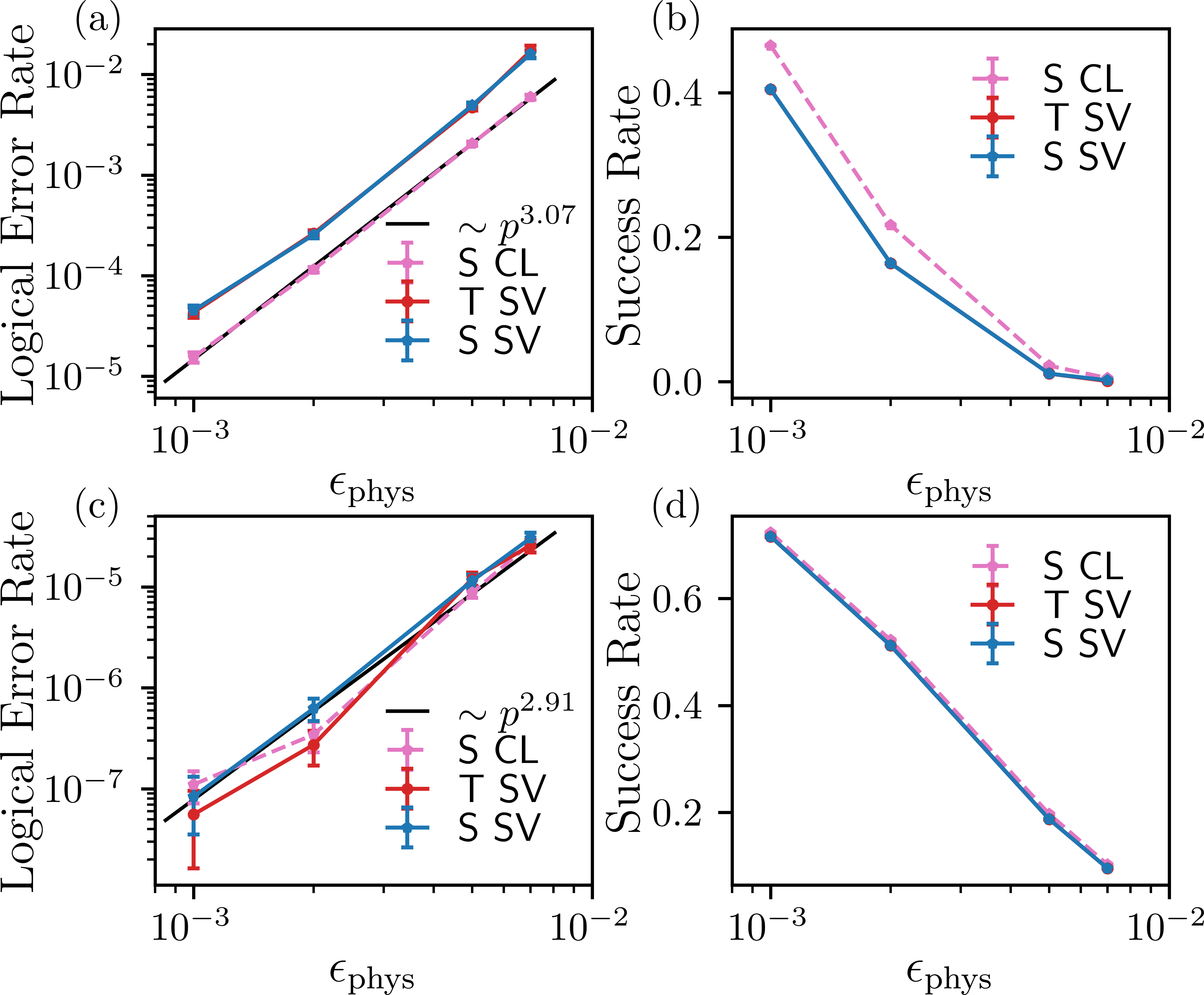}
\caption{\textbf{Performance of magic state preparation using transversal gate.} (a) The logical infidelity and (b) the success rate of $T\ket{+}_L$ and $S\ket{+}_L$ preparations using transversal $T$ are estimated by state vector (SV) simulation (red line: $T\ket{+}_L$ and blue line: $T\ket{+}_L$) and Clifford (CL) gate simulation (purple line: $S\ket{+}_L$) with varying noise levels. Note that the Clifford circuit simulation is simulated under the uniform noise model, while the state vector simulation is executed under an approximate uniform noise model, which serves as an upper bound for the uniform noise model and can be simulated using fewer qubits. The exponents of fitting lines are calculated using $\mathbf{Scipy.state.linregress}$ function. Error bars correspond to the standard error of the mean.}
\label{fig:Fig2_transversal}
\end{figure}

\begin{figure*}[t]
\centering
\includegraphics[width=1\linewidth]{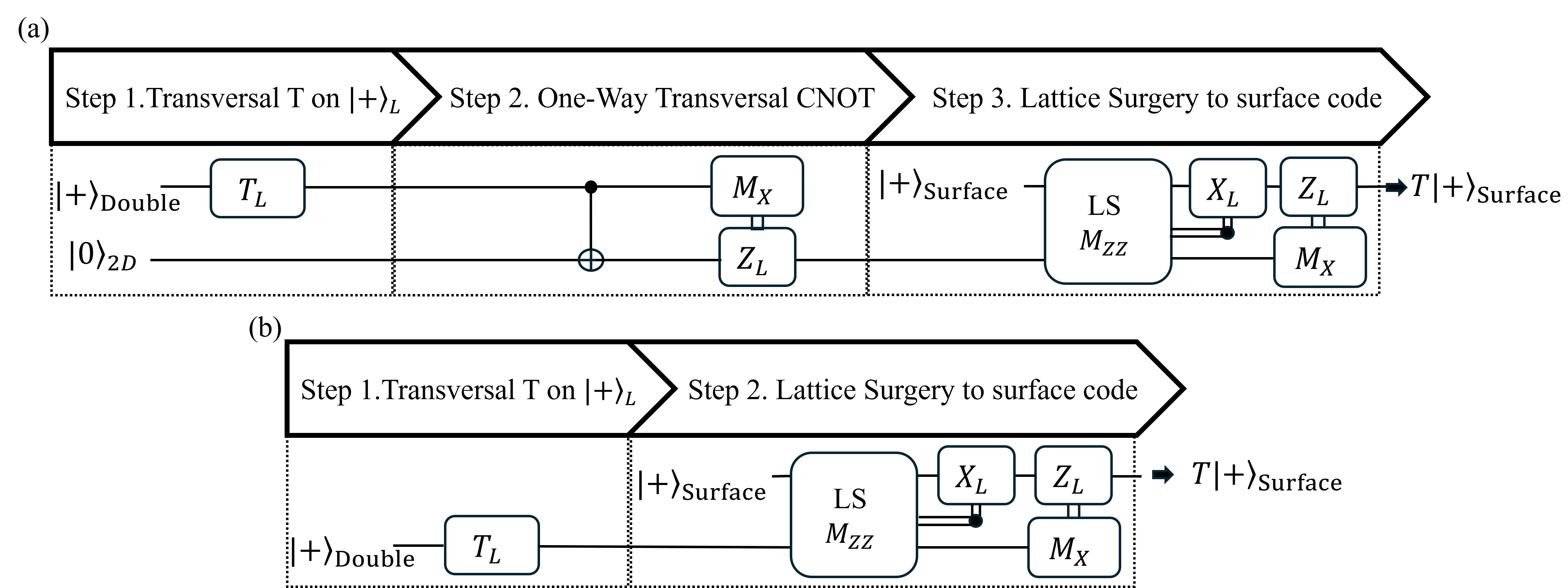}
\caption{\textbf{The workflow of the end-to-end process for preparing a magic state using the doubled color code.} The procedure of end-to-end magic state preparation using transversal gate and lattice surgery from (a) $2$D color code and (b) double color code into a large rotated surface code is depicted. At the first step, before the transversal $T$ gate (shown as $T_L$), one round of $Z$-type stabilizer measurements is applied. For the second step at (a), CS is applied from doubled color code to $2$D color code using a one-way transversal $\overline{\text{CNOT}}$ gate. Then, the logical $X$ is measured (represented by $M_X$). If the outcome gives $1_L$, the logical $Z$ will be applied to the $2$D color code. Otherwise, no operation is implemented on the $2$D color code. In the final step of (a) and (b), the color code is converted into a rotated surface code using lattice surgery (LS) that also measures the $ZZ_L$ (shown as $M_{ZZ}$) of the surface and color codes. If the $M_{ZZ}$ measurement gives a non-logical zero, $X_L$ operations are applied to the surface code. Finally, the color code is measured in the $X_L$ basis. If the measurement outcome is not logical zero, $Z_L$ gates are implemented on the surface code.}
\label{fig:workflow}
\end{figure*}

Magic State Distillation (MSD) ~\cite{Bravyi2005,Bravyi2012,Haah2018,Litinski2019} is one of the popular and explored techniques to fault tolerantly prepare magic states. However, MSD's huge space-time overhead is a bottleneck for near term implementation. Recently, magic state cultivation (MSC) \cite{Gidney2024} was developed to obtain a good enough magic state with a cost equivalent to performing a lattice surgery based CNOT on a surface code. Un-grown MSC has recently been experimentally demonstrated~\cite {Rosenfeld2025} on a superconducting platform. All MSC protocols transfer high-quality magic states to the surface code, which can be easily decoded using an MWPM decoder. $T\ket{+}_L$ cultivation on the color code can be achieved through code grafting~\cite{Gidney2024} or lattice surgery onto the surface code~\cite{Hirano2025}. Moreover, $T\ket{+}_L$ cultivation on $\mathrm{SRP}$-$3$ $(5)$ codes can be morphed into an $\mathrm{RP}^2$ code, a variant of the surface code~\cite{Chen2025}. In this context, the fold-transversal $H_{XY}$, which is composed of a transversal X gate and a fold-transversal $S$ gate on the unrotated surface code, enables one to cultivate magic states on the surface code~\cite{Sahay2025}. Since simulating non-Clifford gates on a classical computer is computationally expensive, most simulations rely on the assumption that the discrepancy between $T\ket{+}_L$ (non-Clifford) and $S\ket{+}_L$ (Clifford) cultivation does not vary with code distance or noise strength, thereby using $S\ket{+}_L$ cultivation as a proxy for $T\ket{+}_L$ cultivation. However, it is not clear that this assumption is warranted. For instance, recent simulations for ungrown MSC on the $d=5$ color code show that $T\ket{+}_L$ cultivation infidelity is $8$ times that of $S\ket{+}_L$ cultivation~\cite{Li2025}. Moreover, a recent paper shows that the logical infidelity discrepancy between proxy state and real magic state cultivation becomes larger and larger as the noise strength decreases~\cite{chase2026}. %Here, by growing, we mean expanding a code region so that its support increases while preserving the encoded logical information.

 Although a comparison between MSC and CS protocols was made in Ref.~\cite{Daguerre25}, the authors ignore idling noise for the CS protocol. Moreover, the results do not include the end-to-end simulation.
Here, we re-evaluate the performance of magic state preparation using a transversal non-Clifford gate and CS protocol under idling noise. %[ATS]
Our protocol consists of three primary steps: first, we prepare a high-fidelity magic state on a small-distance QEC code utilizing a transversal non-Clifford gate. Second, we code-switch this state to an identical-distance $2$D color code. Finally, the $2$D color code is grown and escaped into a larger-sized surface code. 

We also compare the performance of $T\ket{+}_L$ and $S\ket{+}_L$'s preparation using the CS protocol with and without idling noise. We discuss the code growing strategy and extend the protocol to code distance $d=5$. To reduce space overhead, inspired by Ref.~\cite{Hirano2025}, we establish an LS protocol between a doubled color code and a rotated surface code of which we control the target size. Finally, we propose two protocols composed of a double-phase kickback, with MSC as one component, and a transversal non-Clifford gate to prepare the magic state with higher fidelity.   
% Briefly talk about cultivation on color code varing with noise and show their performance T state cultivation and S state's w/o and w/ idling noise. 

%According to the results, further analysis and numerical evidence for $d=5$ are needed. 

\section{Code Switch and Growing\label{sec2}} % Add pic for workflow.
Here, we study magic state preparation on the doubled color code which has transversal $T$ gate. The doubled color code is particularly suitable for near-term devices with all-to-all connectivity and long lifetime physical qubits such as neutral atom~\cite{Bluvstein2025} and trapped ion systems~\cite {Daguerre2025} due to their low idling error rates. To test the consistency of performance between $T\ket{+}_L$ and $S\ket{+}_L$ preparation, we simulate both states' preparation under a uniform noise model with and without idling noise on a state vector simulator. To mitigate the effect of idling noise, we shorten the circuit depth for $\ket{+}_L$ preparation using the MQT QECC package~\cite{mqt,Peham2025} and implement stabilizer measurement using Bell states. However, it is computationally expensive to do a state vector simulation of this circuit, due to the large number of qubits (more than 32) involved. Thus, we approximate the corresponding noisy circuit (more details about the approximate noisy circuit in Appendix.~\ref{sec:approx_cir}) with a smaller, noisy circuit. %This simulation also helps us test the consistency of performance between the $T\ket{+}_L$ preparation and the $S\ket{+}_L$. %[ATS - PLEASE COMPLETE THIS DISCUSSION]. %We also simulate the original circuit for $S\ket{+}_L$ preparation using Clifford circuit simulation to verify this bond.

In Fig.~\ref{fig:Fig2_transversal}, we observe that both state preparations using transversal gates have consistent performance between the $S\ket{+}_L$ and the $T\ket{+}_L$ preparation on the [[15,1,3]] tetrahedral code. In the upper panel of Fig.~\ref{fig:Fig2_transversal}, although the Clifford circuit simulation under  idling noise yields different results from the state vector simulation, the state vector simulation result shows the actual outcome when a magic state is prepared using the transversal gate. The logical error rate achieved is about $1.3\times10^{-5}$ with a physical gate error rate of $10^{-3}$, and corresponding success rate around $46\%$. The corresponding $S\ket{+}_L$ preparation's logical infidelity physical error rate  exponent of roughly $3.07$, which is close to the $[[15,1,3]]$ color code distance. Without idling noise, the result of the Clifford simulation matches that of the state vector simulation. The corresponding infidelity can be below $10^{-7}$, with a success rate of about $ 65$ percent, when the physical gates' noise is around $10^{-3}$. Based on the consistency of $S\ket{+}_L$ and $T\ket{+}_L$'s logical infidelity from the approximate noisy circuit, we believe that $T\ket{+}_L$ preparation on $[[15,1,3]]$ code has a similar performance as $S\ket{+}_L$'s even in the state vector simulation. 
% Show the infidelity of T|+> and S|+>  using transversal gate on [[15,1,3]] code and T|+> and S|+> code switch for [[10,1,2]] and [[7,1,3]]
% Describe the code switch protocol in Daguerre's paper

Although it is hard to expand this doubled color code into a color code with a higher code distance, one can alternatively employ a CS protocol for magic state preparation. The CS protocol~\cite{Daguerre25} for magic state preparation shares the same fundamental ideas as magic state cultivation: preparing a high-fidelity magic state on a small code with full postselection and then immediately grow the code into a surface code with a larger code distance. In Fig.~\ref{fig:workflow}(a), we depict the protocol:   
\begin{enumerate}
    \item Prepare the magic state in the $3$D color code by applying the transversal $T$ gate on $\ket{+}_L$ state. Before the transversal $T$ gate, $Z$-type stabilizers are measured for postselection.
    \item Use the transversal $\overline{\text{CNOT}}$-based CS proposed in Ref.~\cite{Heussen2025} to teleport the state into the $2$D color code, which initially is prepared in the state $\ket{0}_L$.
    \item Grow the color code into a grafted surface code proposed by the Google team~\cite{Gidney2024} or into a rotated surface code via lattice surgery~\cite{Hirano2025}.
    
\end{enumerate}
This method requires a step to code switch the $3$D color code to the $2$D color code. Alternatively, in step 2, we directly convert the $3$D color code into a rotated surface code via LS, shown in Fig.~\ref{fig:workflow}(b). Thus, the protocol becomes
\begin{enumerate}
    \item Prepare the magic state in $3$D color code by applying the transversal $T$ gate on the $\ket{+}_L$ state. Before the transversal $T$ gate, the $Z$-type stabilizers are measured for postselection.
    \item Directly expand the code into a larger rotated surface code via LS.
    
\end{enumerate}
In the subsequent section, we will present examples of different code distance CS protocols and details on growing using LS.

\subsection{$d=2$ case} % Introduce [[10,1,2]] code vs [[7,1,3]] 
The smallest code that supports the transversal $T$ gate is the [[10,1,2]] Vasmer-Kubica code
%by applying the physical $T$, $T^{\dagger}$, and $CCZ$ gates
~\cite{Vasmer2022} (more details in Appendix). Although CS between a [[10,1,2]] code and a [[7,1,3]] code has been proposed and demonstrated on trapped-ion quantum hardware~\cite{Pogorelov2025}, the protocol requires additional stabilizer measurements, which are time-consuming, and therefore prone to errors. Alternatively, based on Ref.~\cite{Heussen2025}, one can teleport the state's information from [[10,1,2]] code to [[7,1,3]] code using a one-way transversal CNOT gate connecting the bottom face of the [[10,1,2]] code to the [[7,1,3]] code, thus avoiding stabilizer measurements. This is possible since the stabilizer group and the logical Pauli operators of the two codes remain invariant after the transversal $\overline{\text{CNOT}}$ gate
\begin{align}\label{eq:t_o_2}
    \overline{\text{CNOT}}\bar{X}_{[[10,1,2]]}\overline{\text{CNOT}} &= \bar{X}_{[[10,1,2]]}\bar{X}_{[[7,1,3]]}\\
    \overline{\text{CNOT}}\bar{Z}_{[[7,1,3]]}\overline{\text{CNOT}} &= \bar{Z}_{[[10,1,2]]}\bar{Z}_{[[7,1,3]]} 
\end{align}
where $\bar{X}_{[[10,1,2]]}$ ($\bar{X}_{[[7,1,3]]}$) and $\bar{Z}_{[[10,1,2]]}$ ($\bar{Z}_{[[7,1,3]]}$) are respectively $X$ and $Z$ logical operators for the $[[10,1,2]]$ ($[[7,1,3]]$) code. Thus, one can also apply the CS protocol to prepare $T\ket{+}_L$ on a 2D color code.

After CS to the $d=3$ Steane code, one also measures the Vasmer-Kubica code in the $X_L$ basis to determine whether the logical $Z$ operator is applied to the Steane code or not. The logical $X$ operator of the Vasmer-Kubica code is 
\begin{align}\label{eq:X_L}
     X_L=X_1X_2X_3X_4X_5X_6X_7 
\end{align}
where the subscript labels of $X$ correspond to the position shown in Fig.~\ref{fig:15_1_3}(a) in the Appendix. Hence, only $7$ qubits at the bottom are measured in the $X$ basis for $X_L$ measurement results. However, one can still measure all $10$ qubits to obtain other $ X$-stabilizer results. If any stabilizer outcomes give a non-trivial syndrome, one discards the corresponding shot. However, the $Z$ type error's code distance is only $2$. Although the code can be switched to a $d=3$ $2$D color code, the $2$-weight Z type errors will dominate the logical error. Hence, we focus on higher distance codes instead.

\subsection{$d=3$ case} % [[15,1,3]] code vs [[7,1,3]]

\begin{figure}
    \centering
    \includegraphics[width=\linewidth]{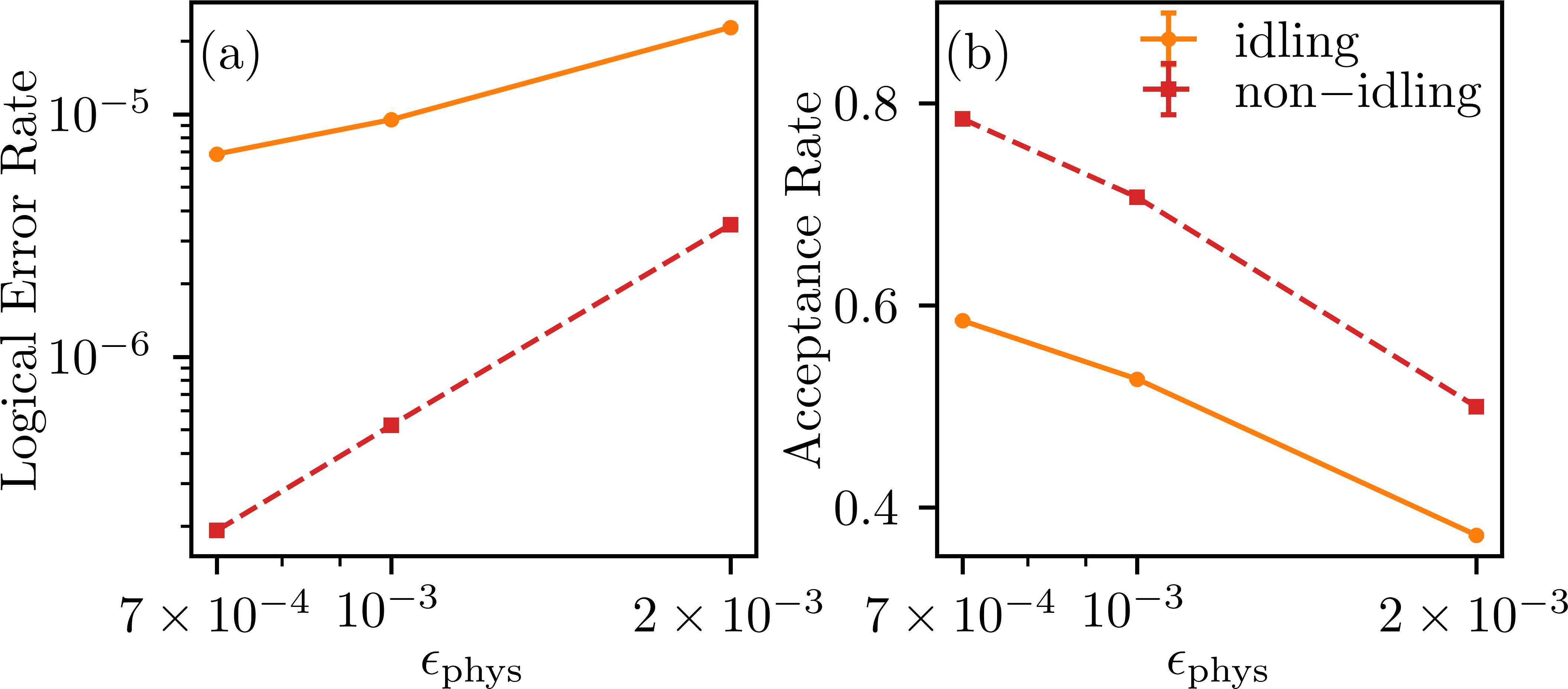}
    \caption{\textbf{The performance for ungrown $d=3$ code switch protocol (with and without the idling noise).} (a) The logical error rates and (b) the acceptance rates varying with the physical noise strength for $[[15,1,3]]$ (with idling noise in orange (`$\cdot$' markers) and with idling noise labeled in red dashed lines (`$\square$' marker).) CS protocol. The number of rounds is $r=2$ after the transversal CNOT code switch. The error bar is the standard error from the successful shots (errors bars are not visible in the plot as the errors bars are order of magnitude smaller than the values of lagical errors and acceptance rate).}
    \label{fig:ungrown_d3}
\end{figure}
For $d=3$, the $3$D color code switch to the $d=3$ Steane code via transversal $\overline{\text{CNOT}}$ gate has been demonstrated in Ref.~\cite{Daguerre25}. Although one can measure all stabilizers with only four extra ancilla qubits using the method in Ref.~\cite{Daguerre25}, unparallelized stabilizer measurements under idling noise end up inducing more logical error. Here, inspired by Ref.~\cite{Takada2024},  we implement a faster stabilizer measurement using Bell pairs to measure all $Z$ ($X$)-type stabilizers in parallel. For the $[[15,1,3]]$ code, a total of $20$ extra qubits are required to measure all $10$ $Z$ stabilizers. Measuring $4$ $X$ stabilizers requires a total of $8$ extra qubits. After CS, we also use Bell pairs to optimize the parallelization of the Steane code's syndrome measurement. The Bell pairs help us obtain $Z$ stabilizer measurement outcomes with a circuit depth of $8$. Moreover, in order to optimize the performance of [[15,1,3]] code preparation in $\ket{+}_L$, we implement the MQT QECC package~\cite{mqt,Peham2025}, which allows us to prepare the $\ket{+}_L$ with circuit depth $5$ and an extra $8$ flag qubits to verify the tetrahedral code $\ket{+}_L$. In total, the space-time volume, which is defined as the number of active qubits times the circuit depth, for initializing $\ket{+}_L$ is $214$ (more details in the Appendix). For the initialization of the $d=3$ Steane code, we implement the same $\ket{+}_L$ preparation method as Ref.~\cite{Bluvstein2025}, which prepares $\ket{0}_L$ on the $d=3$ Steane code with circuit depth $4$, using $2$ flag qubits to check the quality of $\ket{+}_L$. For the entire $d=3$ $CS$, the total space-time volume is $615$; around $50$ percent less than that in Ref.~\cite{Daguerre25} (more details in Appendix).       

To test its performance, we also simulate the CS under uniform noise models with and without idling noise. However, simulating $T\ket{+}_L$ preparation between $d=3$ Steane code and $3$D color code involves at least $22$ data qubits, which does not include the ancilla qubits for syndrome measurement. The corresponding state vector simulation with at least $10^{8}$ shots is complicated. Thus, similarly to Ref.~\cite{Gidney2024}, we simulate $S\ket{+}_L$ preparation using Clifford circuit simulation instead. To evaluate the logical infidelity of the magic state $S\ket{+}_L$ prepared using CS, we add one extra ideal syndrome measurement after a few rounds of noisy measurements in order to filter out all of the residual errors. We also apply the ideal logical $S^\dagger$ gate after the ideal syndrome measurement. We measure the entire $2$D color code's data qubits in the $X$ basis. We calculate the corresponding logical error rate as the number of $1_L$ results divided by the total number of successful shots that yield all-zero syndrome measurements. 

In Fig.~\ref{fig:ungrown_d3}, the performance of $d=3$ CS is visualized. With idling noise, the logical error rates are around $10^{-5}$ when physical gates' error rates range from $2 \times 10^{-3}$ to $7 \times 10^{-4}$. Compared to the performance of ungrown $T\ket{+}_L$ cultivation in $d=3$ Steane code~\cite{Gidney2024}, the performance of $d=3$ CS is around $10$ times worse than MSC. On the other hand, the success rates are around $55\%-60\%$, which is close to the success rate of ungrown MSC. Without idling noise, when physical gates' error rate is at $10^{-3}$, the logical error rate is about $5 \times 10^{-7}$, which is close to the result from Ref.~\cite{Daguerre25}. However, since we implement an extra three rounds of syndrome measurement after CS, the success rate is $70\%$, which is lower than the $84\%$ of Ref.~\cite{Daguerre25}.  

\begin{figure}
    \centering
    \includegraphics[width=\linewidth]{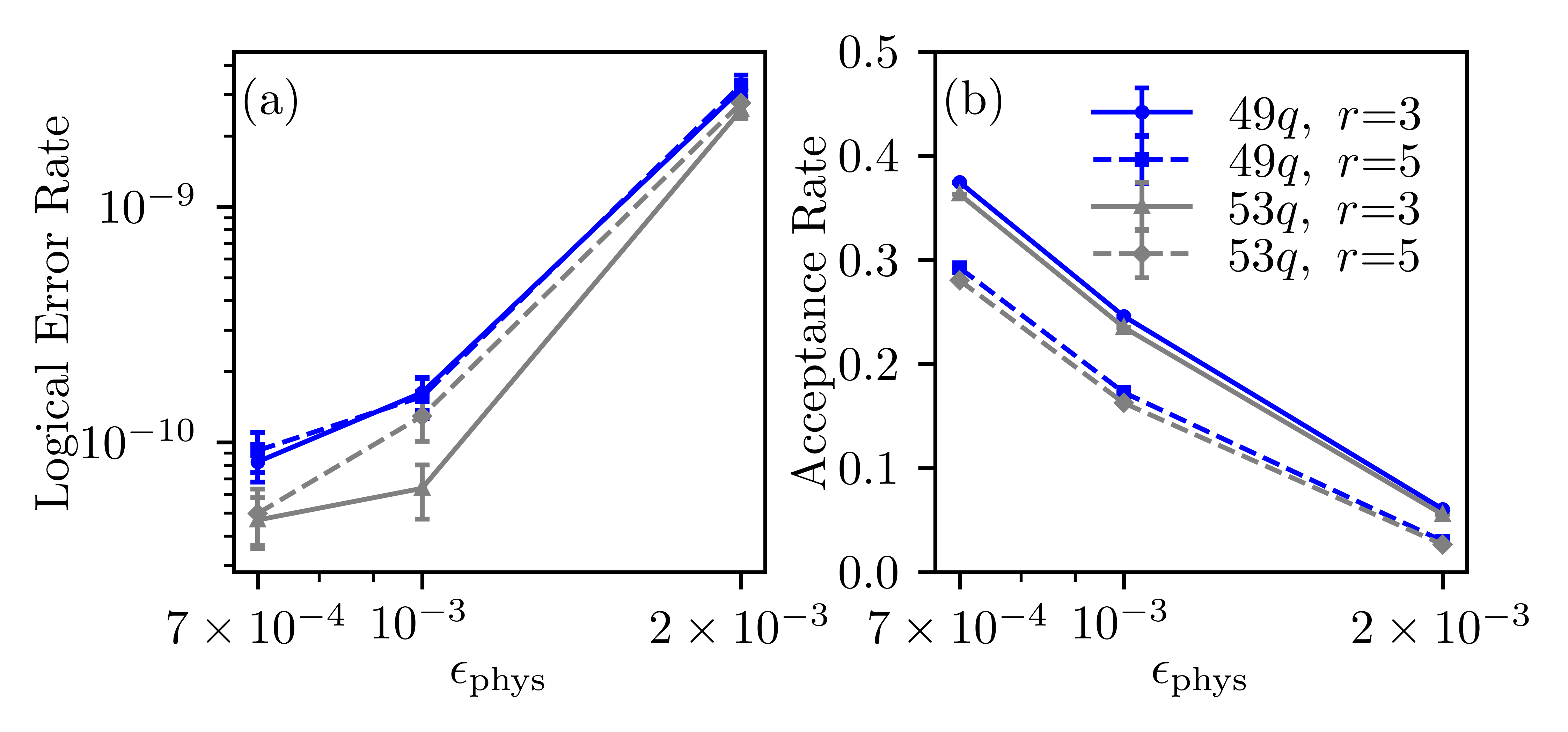}
    \caption{\textbf{The performance for ungrown $d=5$ code switch protocol (without idling noise).} (a) The logical error rates and (b) the acceptance rates varying with the physical noise strength for $[[49,1,5]]$ ($49q$ in blue lines) and $[[53,1,5]]$ ($53q$ in green lines) CS protocol. The dashed line and solid lines means different number of syndrome measurement rounds ($r=5$ and $r=3$ respectively) after the transversal CNOT code switch. The error bar is the standard error from the successful shots.}
    \label{fig:ungrown_d5}
\end{figure}

\begin{figure*}[t]
\centering
\includegraphics[width=1\linewidth]{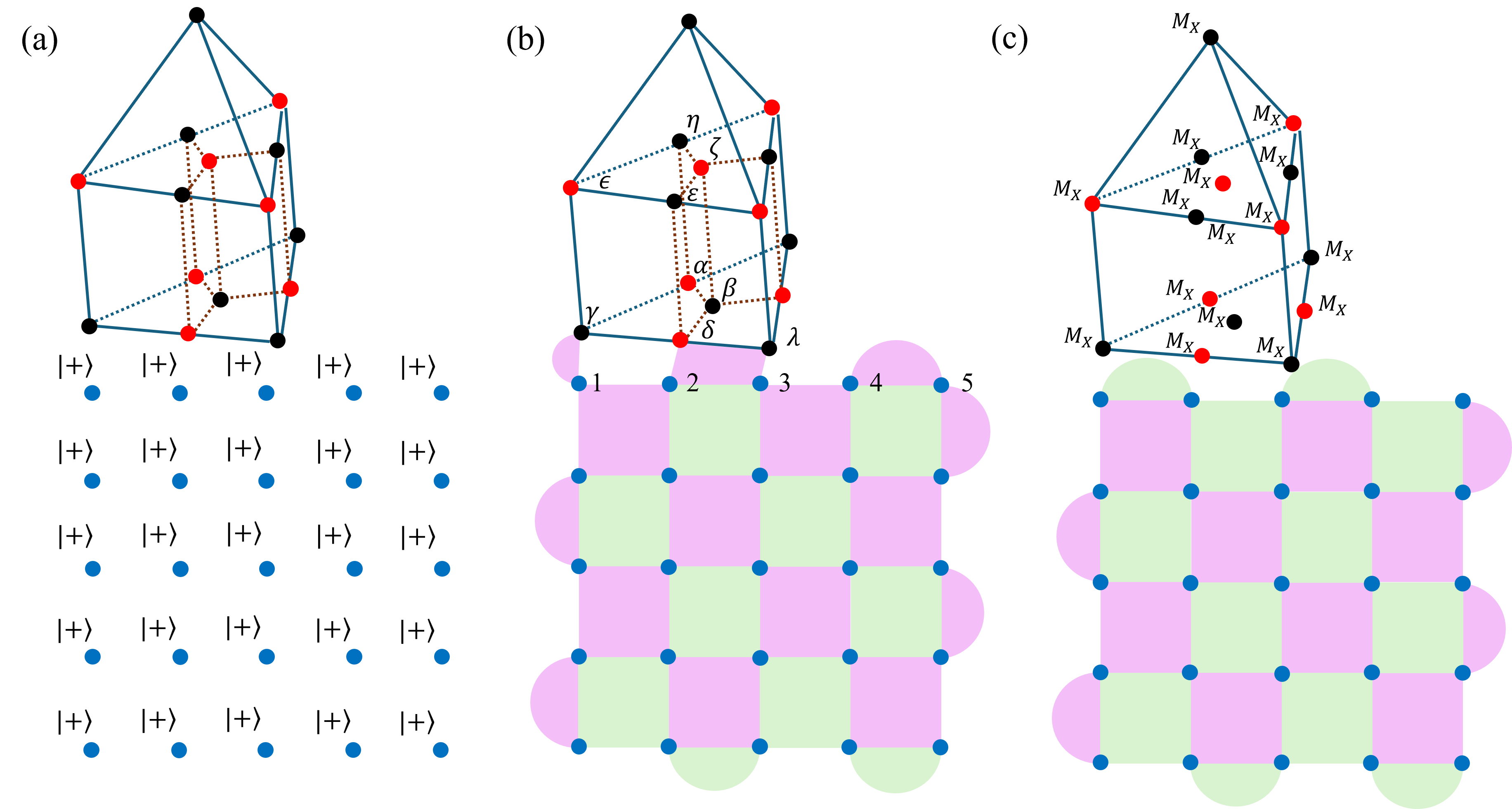}
\caption{\textbf{The illustration of lattice surgery between $3$D color code and rotated surface code.} (a)The surface code initialized with all data qubits $\ket{+}$ states and the color code with $T\ket{+}_L$. (b) The X-type ( green) and Z-type (pink) stabilizers from the surface code and the boundary between surface code and color code. (c) The stabilizers of the surface code after the lattice surgery and $X_L$ measurement of the color code. Since the stabilizer $X_\alpha X_\beta X_\gamma X_\delta X_\epsilon X_\varepsilon X_\zeta X_\eta$ from color code doesn't commute with $Z_\gamma Z_1$ and $Z_\delta Z_\lambda Z_2 Z_3$, the stabilizer becomes $X_1X_2X_\alpha X_\beta X_\gamma X_\delta X_\epsilon X_\varepsilon X_\zeta X_\eta$ during the lattice surgery. After the lattice surgery, the color code is measured in the $X_L$ basis by measuring all qubits individually in $X$ basis (shown as $M_X$). If there are odd number of $1$s result, the logical outcome $1_L$; otherwise, the logical outcome is $0_L$.   }
\label{fig:3D_LS}
\end{figure*}

\subsection{$d=5$ case} % [[49,1,5]] code vs [[17,1,5]] code and [[53,1,5]] code vs [[19,1,5]] code
To extend the doubled color code from $d=3$ to $d=5$, one can construct the 3D $d=5$ color code using a doubling transformation from the $d=5$ 2D color code~\cite{Bravyi2015,Jain2025}. If one chooses a $d=5$ $(4,8,8)$ 2D color code [[17,1,5]], after the doubling construction, the 3D color code becomes a [[49,1,5]] code. On the other hand, if the chosen code is the $(6,6,6)$ 2D color code [[19,1,5]], the corresponding 3D color code is [[53,1,5]] using the doubling construction. The constructed $3$D color codes [[49,1,5]] ([[53,1,5]]) also support the transversal $\overline{\text{CNOT}}$ CS with the [[17,1,5]] ([[19,1,5]]) Steane code~\cite{Sulliva2024}. These 3D color codes can be visualized as recursively stacked color codes~\cite{Jochym-O'Connor16,Tansuwannont2022}. As Fig.~\ref{fig:d5_3S_code} shows, the corresponding code resembles a qubit on top of stacks of paired layers of $d=3$ 2D color codes and paired layers of $d=5$ 2D color codes. Moreover, the tetrahedral color code can be extended from $d=3$ to $d=5$~\cite{Butt2025}. However, the corresponding $d=5$ tetrahedral color code requires more physical qubits for encoding ($65$ qubits for $d=5$). Thus, we only focus on the recursively stacked color code for that case.

In order to fault-tolerantly prepare $\ket{+}_L$ and $\ket{0}_L$ for double and $2$D color codes, respectively, we employ the MQT QECC package~\cite{mqt,Peham2025} with extra flag qubits~\cite{Forlivesi2025} to check whether the prepared state is correct or not. Before the end-to-end CS-based magic state preparation, we simulate ungrown magic state preparation to understand how many rounds of syndrome measurements are required after CS. However, the simulation of $T\ket{+}_L$ preparation using a transversal $T$ gate is beyond the Clifford simulation as the $T$ gate is non-Clifford. Thus, we simulate $S\ket{+}_L$ proxy state preparation instead and estimate the performance of the $T\ket{+}_L$ preparation based on the $d=3$ state vector simulation result. Moreover, since the $d=5$ double color code has $35$ Z stabilizers for $[[49,1,5]]$ code ($38$ Z stabilizers for $[[53,1,5]]$ code), with idling noise, the corresponding success rate to get all zero syndrome measurements is very low. Hence, we only simulate $d=5$ CS under uniform noise without the idling noise.   

In Fig.~\ref{fig:ungrown_d5}(a), the logical error rates of the preparation of the magic state using CS are around $10^{-11}-10^{-9}$ with noise strengths in the range $7 \times 10^{-4}-2\times10^{-3}$. Two extra rounds of syndrome measurement do not improve infidelity. Instead, as Fig.~\ref{fig:ungrown_d5}(b) shows, the corresponding success rates are about $20$ percent less than those with fewer rounds of syndrome measurement. Overall, the success rate for both codes is about $10\%-40\%$ when the physical error rate is around $2 \times 10^{-3}-7\times10^{-4}$. While $[[53,1,5]]$ slightly outperforms $[[49,1,5]]$ in terms of infidelity, $[[49,1,5]]$'s CS has a higher success rate than $[[53,1,5]]$'s.

\subsection{Code Growing}

To maintain low logical errors and perform error correction on the code, it is necessary to expand the small code into a larger code immediately. However, directly expanding the color code into a larger color code introduces too many logical errors due to the Chromobius decoder's poor performance~\cite {Gidney2024,Gidney2022}. Alternatively, one can expand the color code into a code that can be decoded with minimum weight perfect matching (MWPM) decoder~\cite{Higgott2025}. One can expand the color code using grafting techniques into a surface code. Another way is to switch the $2$D color code to a larger surface code via lattice surgery. Here, we not only implement LS from the $2D$ color code to the surface but also develop an LS protocol to transform a doubled color code into a larger surface. After the LS, one needs to apply several rounds of syndrome measurement and simultaneously wait for the decoding result from the decoder. Based on the complementary gap's result calculated from the decoder~\cite{Gidney2023}, one can set up a threshold to discard corrupted results and get a higher quality magic state.    
% Lattice surgery from 2D color code to 2D surface code. (extend to d=5?) growing protocol of Google and lattice surgery from the Japanese group.
% Lattice surgery from 3D color code to 2D surface code. 
%(Grafting)

\begin{figure*}
    \centering
    \includegraphics[width=\linewidth]{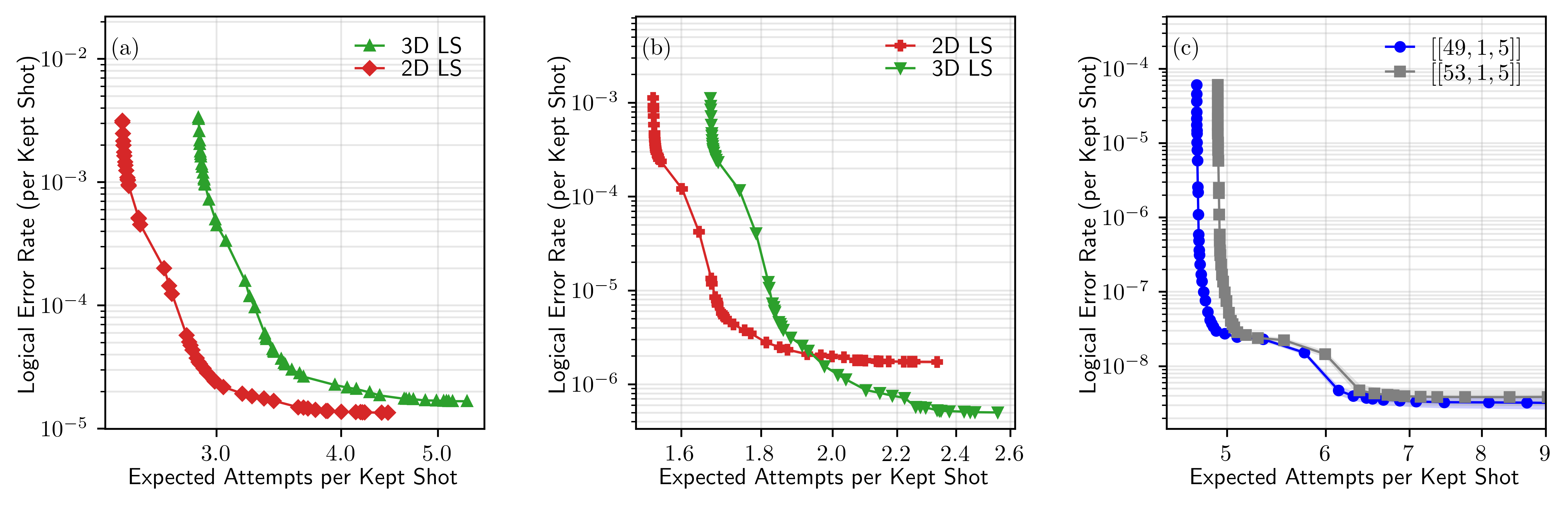}
\caption{\textbf{End-to-end simulation results for magic state preparation using a transversal non-Clifford gate.} The logical infidelity of simulation for (a) $d=3$ with idling noise, (b) $d=3$ without idling noise, and (c) $d=5$ without idling noise. The curves vary with complementary gap threshold to postselect the shot with logical error-prone syndrome. In the $d=3$ cases, $3D$ LS ($2$D LS) means that the end-to-end simulation including the lattice surgery from doubled ($2D$) color code to rotated surface code ($d_f=11$). For $d=5$ case, $[[49,1,5]]$ ($[[53,1,5]]$) represents the end-to-end simulation including the code switching from $[[49,1,5]]$ ($[[53,1,5]]$) code to $[[17,1,5]]$ ($[[19,1,5]]$) code and lattice surgery from $2$D color code to rotated surface code ($d_f=11$). The shaded areas indicate standard sampling error. }
\label{fig:end2end}
\end{figure*}

One of the issues for LS between a $2$D color code and the surface code is the parallelization of the color code and the surface code's stabilizer measurements. The circuit depth of Surface code's X-type and Z-type stabilizers measurement is shorter than those of the $2$D color code using superdense cycle. In order to parallelize color code and surface code's syndrome extraction, differently than Ref.~\cite{Hirano2025}, we implement syndrome measurement using a standard color code cycle with Bell flagging. The X(Z)- type stabilizer measurement of the color code roughly matches one round of the surface code cycle. Hence, for $d=3$ case, there are three rounds of surface code stabilizer measurement, while there are only two rounds for the $2$D color code (including X stabilizer measurement at the final measurement to get a logical X result) during LS. The stabilizers at the boundary between $2$D color code and the rotated surface code are measured $3$ times. Moreover, instead of switching to an intermediate-sized rotated surface code, we directly perform LS to the $2$D color code with the surface code at the targeted size.    

The procedure of $3$D color code switching into the rotated surface code via LS is very similar to that from $2$D color code. The only difference is that the syndrome measurement of the $3$D color code is more complex than that of the $2$D color code, whose corresponding stabilizers can be measured using a superdense cycle or a standard color code cycle. Inspired by Ref.~\cite{Takada2024}, using an extra flag qubit paired with a readout qubit as a Bell pair for each stabilizer measurement (in total, 20 ancilla qubits are required), we can parallelize all $10$ Z stabilizer measurements for the [[15,1,3]] code. Similarly, for $8$-weight X stabilizers, we also perform each stabilizer measurement with a Bell pair. We also parallelize all $4$ X stabilizer measurements. Thus, we can also apply LS between the $d=3$ $3$D color code and the rotated surface code. The corresponding LS procedure, shown as Fig.~\ref{fig:3D_LS}, is as follows:
\begin{enumerate}
    \item The data qubits on the rotated surface are all initialized as $\ket{+}$ after the preparation of $T\ket{+}_L$ on the $3$D color code.

    \item The surface code's stabilizers shown in Fig.~\ref{fig:3D_LS}(b) are measured three times. Meanwhile, X or Z stabilizers of the $3$D color code are measured. At even-numbered cycles of the surface code syndrome measurement, the X stabilizers of the color code are measured except for $X_\alpha X_\beta X_\gamma X_\delta X_\epsilon X_\varepsilon X_\zeta X_\eta$ and $X_1X_2X_\gamma X_\delta$ stabilizers since it becomes $X_1X_2X_\alpha X_\beta X_\gamma X_\delta X_\epsilon X_\varepsilon X_\zeta X_\eta$ after the first round of syndrome measurement. Otherwise, Z stabilizer measurements are implemented. In addition, the Z stabilizers, $Z_\gamma Z_1$, $Z_\delta Z_\lambda Z_2 Z_3$, and $Z_4 Z_5$, at the boundary between the color code and the surface code are also measured in every surface code cycle. The results of these stabilizers give us the $ZZ_L$ result which determines whether the $X_L$ operator is applied after LS.  

    \item All data qubits on the $3$D color code are measured in the $X$ basis. The measurement of the color code's X stabilizers including $X_1X_2X_\alpha X_\beta X_\gamma X_\delta X_\epsilon X_\varepsilon X_\zeta X_\eta$ can also be obtained at the same time. If a measurement gives a non-trivial syndrome outcome, the result is discarded. The $X_L$ measurement is determined by the parity of the total measurement. If it is even, $X_L=0_L$; otherwise, $X_L=1_L$. This also determines whether the $Z_L$ operator is implemented.  
\end{enumerate}
After LS, in order to decode using a correlated MWPM decoder~\cite{Gidney2023,Higgott2025}, we discard shots with non-trivial syndromes from the color code or the boundary between the color code and the surface code. Moreover, similarly to Ref.~\cite{Takada2024}, we use a noise model without noise triggering the stabilizer syndrome for the MWPM.

\section{End-to-End Simulation}
Here, to evaluate the entire CS protocol's performance, we execute end-to-end simulation of the CS protocols. Since simulation beyond the Clifford circuit is expensive, we simulate $S\ket{+}_L$ in place of $T\ket{+}_L$ for the CS protocol. Due to the consistency between $S\ket{+}_L$ and $T\ket{+}_L$ preparation (the result shown in Fig.~\ref{fig:Fig2_transversal}), the result of $S\ket{+}_L$ preparation should be close to that of $T\ket{+}_L$. We additionally assume that the complementary gap computation takes $10$ rounds of surface code stabilizer measurement. After $10$ rounds, we discard corrupted outcomes based on whether the corresponding gap passes the threshold. Since stabilizer measurement on the $d=5$ doubled color code washes out too many shots, we do not simulate $3$D LS from $d=5$ doubled color code to surface code. 

In Fig.~\ref{fig:end2end}, end-to-end simulations for $d=3$ and $d=5$ are visualized. The leftmost data point of each curve represents the logical error rate without complementary-gap-based soft postselection. The higher threshold cutoff on the complementary gap makes the corresponding logical error rates lower until the logical error (roughly) converges to a bound, which is the lowest logical error rate when using complementary-gap-based postselection. As Fig.~\ref{fig:end2end}(a) shows, with idling noise, the $2$D LS and CS protocol requires fewer attempts per retained shot than $3$D LS to reach its best logical fidelity, around $10^{-5}$. Moreover, the logical infidelity of $3D$ LS is around $2 \times 10^{-5}$, which is similar to the logical error rate of $2$D LS. Without idling noise (shown in Fig.~\ref{fig:end2end}(b)), the infidelity of $2$D LS can reach below $2\times10^{-6}$ with only $2$ attempts per retained shot. On the other hand, the logical error using $3$D LS can reach below $5\times10^{-7}$, but it requires $2.4$ attempts per shot. 

Without idling noise, the $2$D and $3$D LS results above imply that the initialization of the $2$D color code and the transversal CNOT between two QEC codes induce extra logical errors. Thus, the $3$D LS procedure outperforms that of the $2$D LS in terms of infidelity. However, when idling noise is involved, the $3$D LS procedure has a slightly worse outcome than that of $2$D LS. In the $d=5$ case shown in Fig.~\ref{fig:end2end}(c), CS from $[[49,1,5]]$ outperforms that from $[[53,1,5]]$ in terms of success rate and logical fidelity. For the $[[53,1,5]]$ CS, the logical infidelity can achieve $4 \times 10^{-9}$ with a success rate of around $13.9$ percent. In contrast, for the $[[49,1,5]]$ CS, the logical error rate can attain $3 \times 10^{-9}$ with a similar success rate.

In addition to end-to-end performance, we also estimated the expected space-time volume per successful shot for the entire magic state preparation process, defined as
\begin{align}\label{eq:stv}
 V=\frac{1}{f_M}\sum^M_{i} f_i V_i
\end{align}
with $f_i$ and $V_i$ the success rate and spacetime volume for $i$th part of the circuit (resp.). $f_M$ is the total success rate of the entire circuit. The corresponding $V$s are estimated in table~\ref{tab:ls_error_rates}. In the $d=3$ case, with idling noise, the $2D$ LS $V$ is $26.5 \%$ lower than $3$D LS with the same final logical error rate, $1.5 \times 10^{-5}$. Compared to the MSC's $V$ (around $50,000$) on the color code~\cite{Gidney2024,Sahay2025}, the $2D$ LS $V$ is lower, but MSC gives a higher fidelity magic state ($2 \times 10^{-6}$). On the other hand, without idling noise, although $V$ of $3$D LS is still higher than $2$D LS, the corresponding final logical error rate of $3$D LS can reach $6 \times 10^{-7}$. In contrast, the logical error rate of $2$D LS can only approach $2 \times 10^{-6}$. In the $d=5$ case, since we do not parallelize the circuit well, the corresponding space volume of end-to-end is very large. We leave the optimization of the spacetime volume for $d=5$ to the future work.    

\begin{table}[h]
\centering
\begin{tabular}{l c c}
\toprule
 & $V$ & Final logical error rate \\
\midrule
2D LS w/ i.d. noise   & 39655 & $1.5 \times 10^{-5}$ \\
2D LS w/o i.d. noise  & 29115 & $2 \times 10^{-6}$   \\
3D LS w/ i.d. noise   & 53989 & $1.5 \times 10^{-5}$ \\
3D LS w/o i.d. noise  & 34209 & $6 \times 10^{-7}$   \\
\bottomrule
\end{tabular}
\caption{\textbf{The expected spacetime volume per successful shot.} The expected spacetime volume per successful shot to achieve a given logical error rate after complementary-gap-based postselection for 2D and 3D lattice surgery (LS) with and without idling (i.d.) noise. More details about expected spacetime volume may be found in Appendix~\ref{sec: sp_tv}.}
\label{tab:ls_error_rates}
\end{table}

% Showing d=3 (d=5) code switch protocol end to end simulation. 
\begin{figure}[t]
\centering
\includegraphics[width=1.0\linewidth]{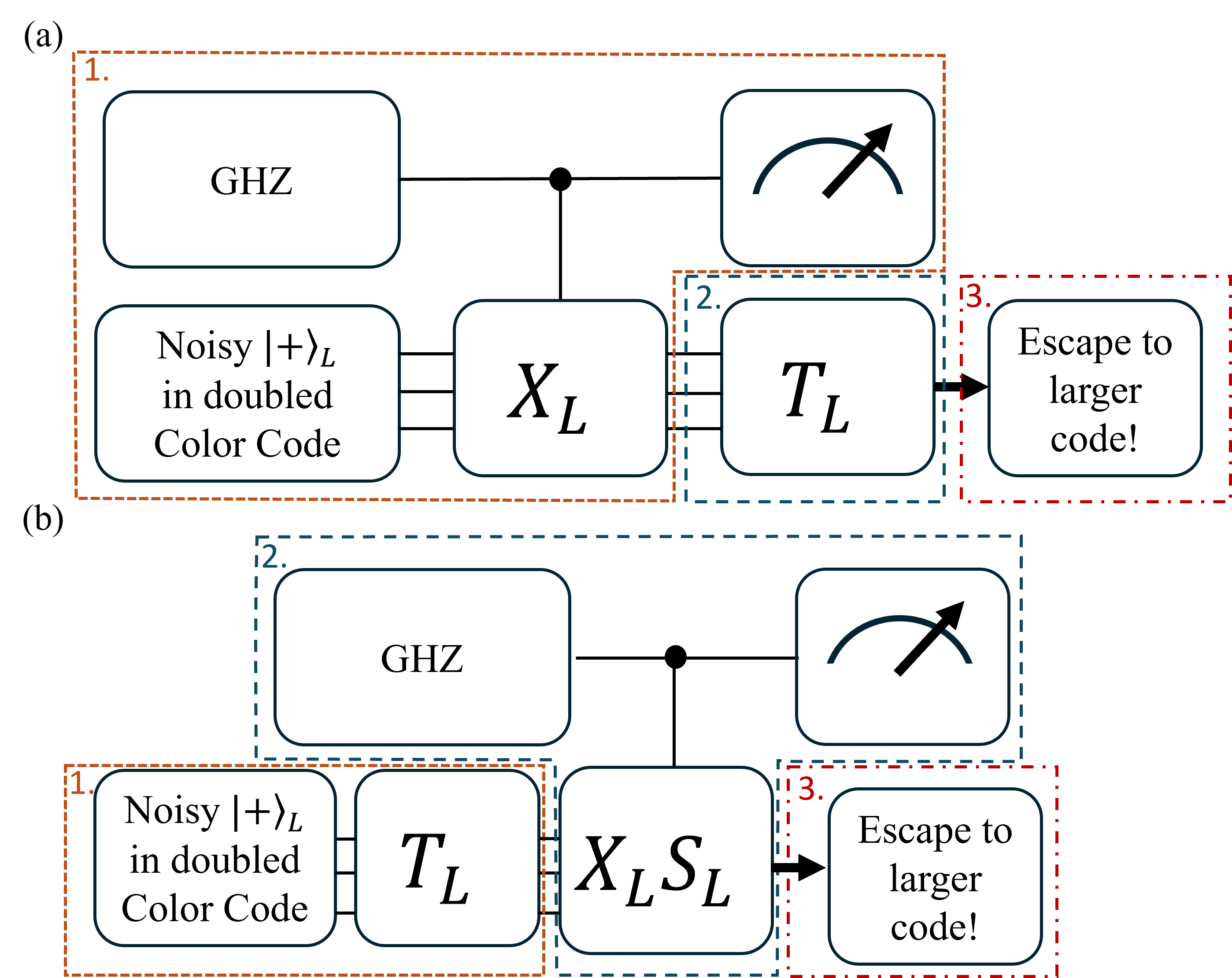}
\caption{\textbf{Illustration of a circuit composed of magic state cultivation and transversal non-Clifford gate.} In the first step, a noisy $\ket{+}_L$ state is prepared. Phase kickback measurement is used to check the quality of $\ket{+}_L$. One disentangles the GHZ state into a product state in the $Z$ basis after the transversal CNOT (represented by control $X_L$ in the figure). If the measurement outcome gives a non-zero result, it indicates that a non-trivial error has occurred in the code. One then directly abandons the shot. After the phase kickback measurement, the second step is to apply the transversal $T$ gate (shown as $T_L$ in the figure) to the $3$D color code. Finally, the double color codes are converted into a large code. One can switch double color code into $2$D color code via a transversal $\overline{\text{CNOT}}$ and transform it into a rotated surface code using lattice surgery. Alternatively, one can directly switch $3$D color code into surface code using lattice surgery.       }
\label{fig:MSC_CS}
\end{figure}

\section{Combination of Cultivation and Transversal Non-Clifford Gate}

% prepare |+>_L with X check and apply transversal T gate? What's advantage? 
Although magic state preparation with transversal non-Clifford gates provides another way to prepare high fidelity magic states, fault-tolerant preparation of 3D codes as $\ket{+}_L$ is challenging, especially for larger codes. Although one can use extra flag qubits to filter out bad logical state preparation, how to implement this is complicated. To address this issue, we propose a new method composed of MSC and code-switching protocols. 

As Fig.~\ref{fig:MSC_CS}(a) shows, our new protocol for preparing the magic states is the following:
\begin{enumerate}
    \item Directly prepare noisy $\ket{+}_L$ (without extra flag qubits) with one round of syndrome measurement then utilize the MSC double phase kickback measurement to check the quality of $\ket{+}_L$. 
    \item Apply transversal $T$ gate on $3$D color code.
    \item Implement CS from 3D color code to 2D color code, and LS to a larger rotated surface code. Alternatively, directly perform LS to a larger rotated surface code from the $3$D color code.
    
\end{enumerate}
Here, based on the transversal $T$ and $S$ gate results shown in Fig.~\ref{fig:Fig2_transversal}, we also assume that the logical noise induced from $T\ket{+}_L$ using this method is similar to or lower than that from $S\ket{+}_L$.

Another protocol combining MSC and transversal non-Clifford gates is to implement the transversal $T$ gate first and apply the phase kickback check subsequently. In Fig.~\ref{fig:MSC_CS}(b), the corresponding protocol is as follows
\begin{enumerate}
    \item Prepare noisy $\ket{+}_L$ (without extra flag qubits) and apply transversal $T$ after preparing the state. After the transversal $T$ gate, 
    \item Implement double phase kickback measurement to check the quality of $T\ket{+}_L$. 
    \item Implement CS from 3D color code to 2D color code, and LS to a larger rotated surface code. Otherwise, directly perform LS to a larger rotated surface code from $3$D color code.
    
\end{enumerate}
Here, the double phase kickback check is similar to Ref.~\cite{Sahay2025} and is composed of transversal $X$ gate and $S$ gate. In this case, one should apply extra physical $T$ gates on the ancilla qubits to cancel the phase generated from $SXT\ket{+}_L =e^{i\frac{\pi}{4}}T\ket{+}_L$. According to Ref.~\cite{Chen2026}, one can apply the $T$ or $T^\dagger$ gates on the auxiliary GHZ state after control $XS^\dagger$ or $XS$. In total, for $d=3$ doubled color code, a $15$-qubit GHZ state is required. Alternatively, control $XS^\dagger$ and $XS$ can be paired together to cancel the corresponding phase on the auxiliary GHZ state. One of the GHZ ancilla qubits receives only one $XS$ control operation. After disentangling the ancilla GHZ state into a Pauli basis product state, $T^\dagger$ is applied on the corresponding ancilla qubit. In this case, a $ 7$-qubit GHZ state is required for the phase kickback check. Similarly, one can also implement $2$ control $XT^\dagger$ and $2$ control $XT$ gates with the same control qubits of GHZ state. Hence, a $ 4$-qubit GHZ state is required.  

\begin{figure}[t]
    \centering
    \includegraphics[width=\linewidth]{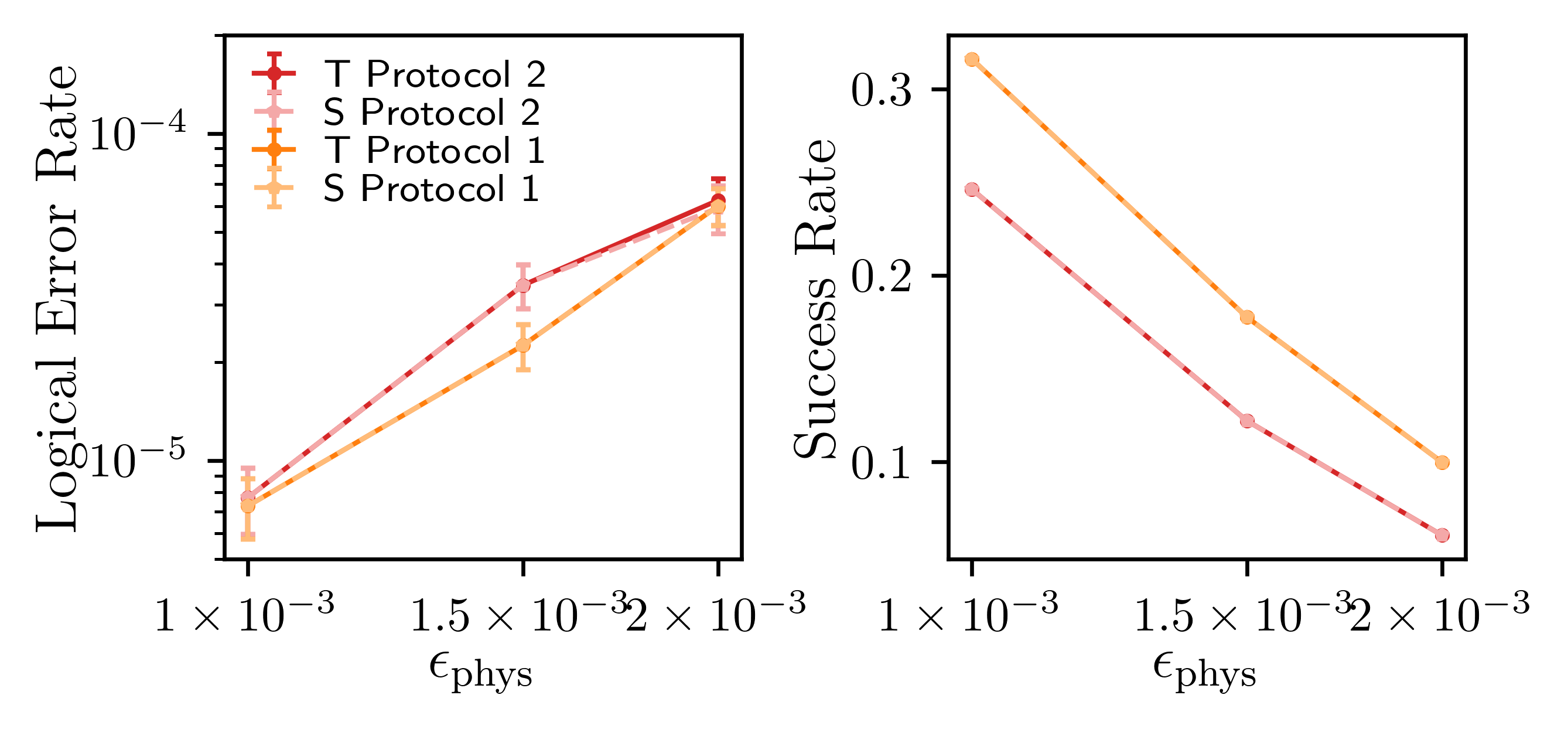}
    \caption{\textbf{The statevector simulation for a combination of cultivation and transversal non-Clifford Gate.} (a) The logical infidelity and (b) the success rate of $T\ket{+}_L$ and $S\ket{+}_L$ preparation using both protocols, which include double phase check to check the quality of $\ket{+}_L$ and $T\ket{+}_L$ or $S\ket{+}_L$. The protocol $1$ ($2$) represents the first (second) protocol, which has a phase kickback double check for $\ket{+}_L$ ($T\ket{+}_L$ or $S\ket{+}_L$). Here, we randomly choose seeds for the simulation, but $T$ and $S$ state's simulation share the same seed.}.
    \label{fig:sv_cult}
\end{figure}

To test the consistency in performance between the preparation $S\ket{+}_L$ and $T\ket{+}_L$ for these two protocols, we implement a state vector simulation before the end-to-end simulation. We apply the same strategy as in the Sec.~\ref{sec2} state vector simulation to estimate the corresponding logical error rate. We choose the $4$-qubit GHZ state for the state-vector simulation to avoid complicated computation. Furthermore, in order to fairly compare the two protocols, we assume that control $XS$ ($XS^{\dagger}$) in the second protocol is a two-qubit gate with a following two-qubit depolarized noise. 

In Fig.~\ref{fig:sv_cult}, the $T\ket{+}_L$ and $S\ket{+}_L$ preparation using both protocols are almost consistent with each other. Compared to the second protocol, the first protocol stands out for its greater logical fidelity and higher success rate. At a physical gate error rate of $10^{-3}$, the logical infidelities of both protocols are below $10^{-5}$, which is much lower than the original result with only a transversal $T$ ($S$) gate (shown in Fig.~\ref{fig:Fig2_transversal}). 
For the end-to-end simulation, we also only simulate $S\ket{+}_L$ for the logical infidelity estimation of both protocols. To optimize the logical infidelity, we use a $ 15$-qubit GHZ state to implement the double-phase-kickback check. Moreover, we use uniform noise with a physical error rate $10^{-3}$ to study how the success rate and logical error rates vary with the complementary gap.     

\begin{figure}[t]
    \centering
    \includegraphics[width=\linewidth]{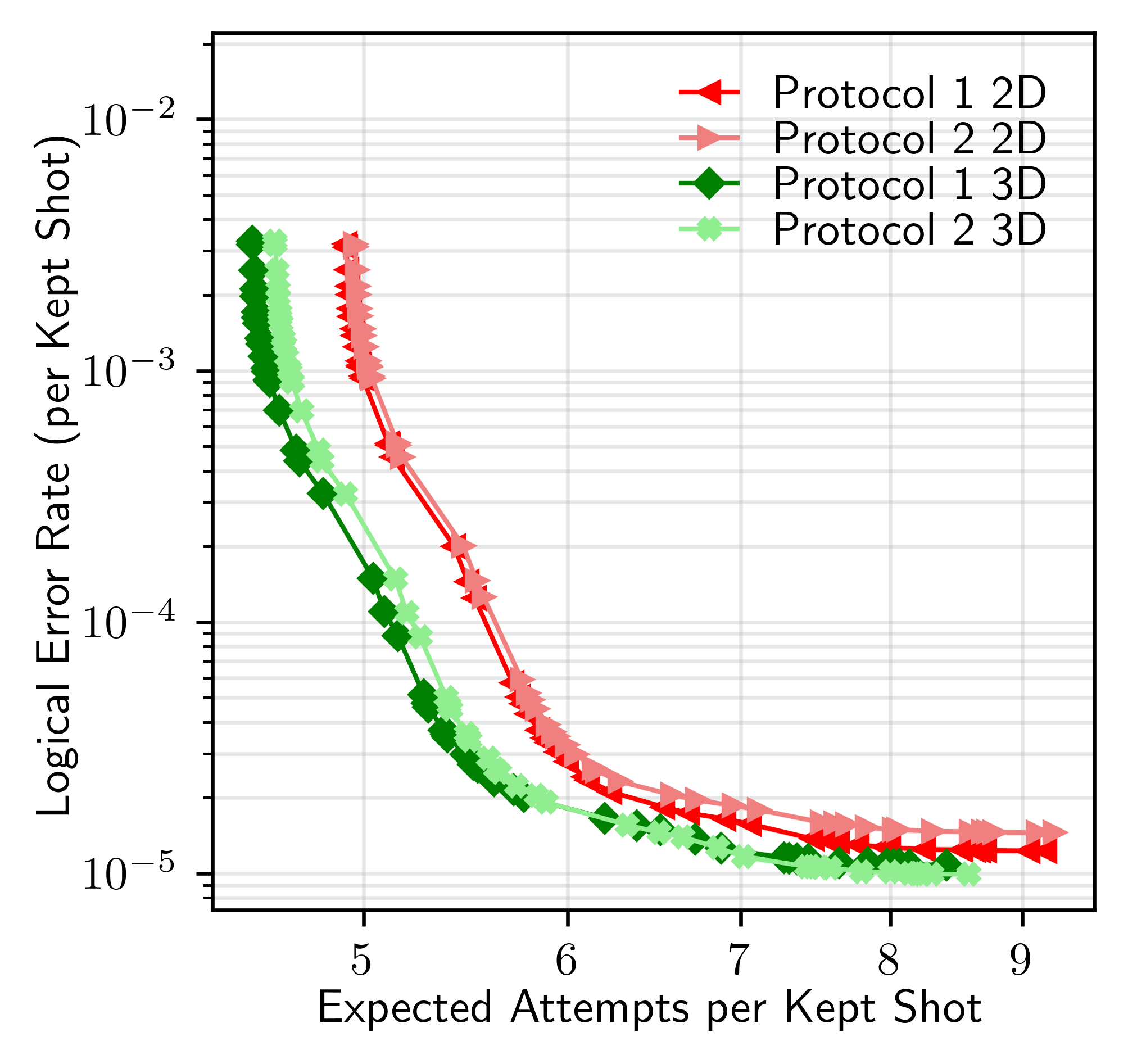}
    \caption{\textbf{The end-to-end simulation for the combination of cultivation and transversal non-Clifford Gate.} The logical infidelity at physical error rate $10^{-3}$ varies with the attempts per kept shot, which is determined by the complementary gap-based postselection. The protocol $1$ (2) refers to the first (second) protocol, which has the double phase kickback check to verify the quality of $\ket{+}_L$ ($S\ket{+}_L$). $2$D ($3$D) means the different escape strategies using lattice surgery between Steane code (double color code) and rotated surface code.}
    \label{fig:GHZ_end2end}
\end{figure}

In Fig.~\ref{fig:GHZ_end2end}, the first protocol with the double-kick-phase check for $\ket{+}_L$ has similar performance as the second protocol with double-kick-phase check for $S\ket{+}_L$. As the complementary gap increases, more and more shots with logical error-prone syndromes get removed. For both protocols, the corresponding logical error rates reach the saturated value, which is around $10^{-5}$ with roughly $8$ attempts per retained shot. Since $3$D LS does not require the preparation of Steane code and transversal CNOT, the corresponding logical error rates of both protocols are lower than $2$D LS. Among these methods, the first protocol with $3$D LS has a slightly better logical error rate than the second protocol. Although the logical infidelity has slightly better performance than the original results ($2$D and $3$D LS with idling noise) from Fig.~\ref{fig:end2end}(a), the expected attempts per retained shot are much higher than the original results.   

In Table~\ref{tab:ls_error_rates_transversal}, the expected spacetime volumes per successful shot for different protocols are estimated. Overall, the first protocol has slightly less expected spacetime volume per successful shot relative to the second protocol, obtaining a logical error rate of $1.5 \times 10^{-5}$. Compared with the original $3$D LS result in Table~\ref{tab:ls_error_rates}, the first protocol and the second protocol with $3$D LS have lower $V$, since the double phase kickback check can filter out more shots before LS expansion. On the other hand, $V$ for the first and second protocols with LS does not outperform that of the original $2$D LS.

\begin{figure}[t]
\centering
\includegraphics[width=0.8\linewidth]{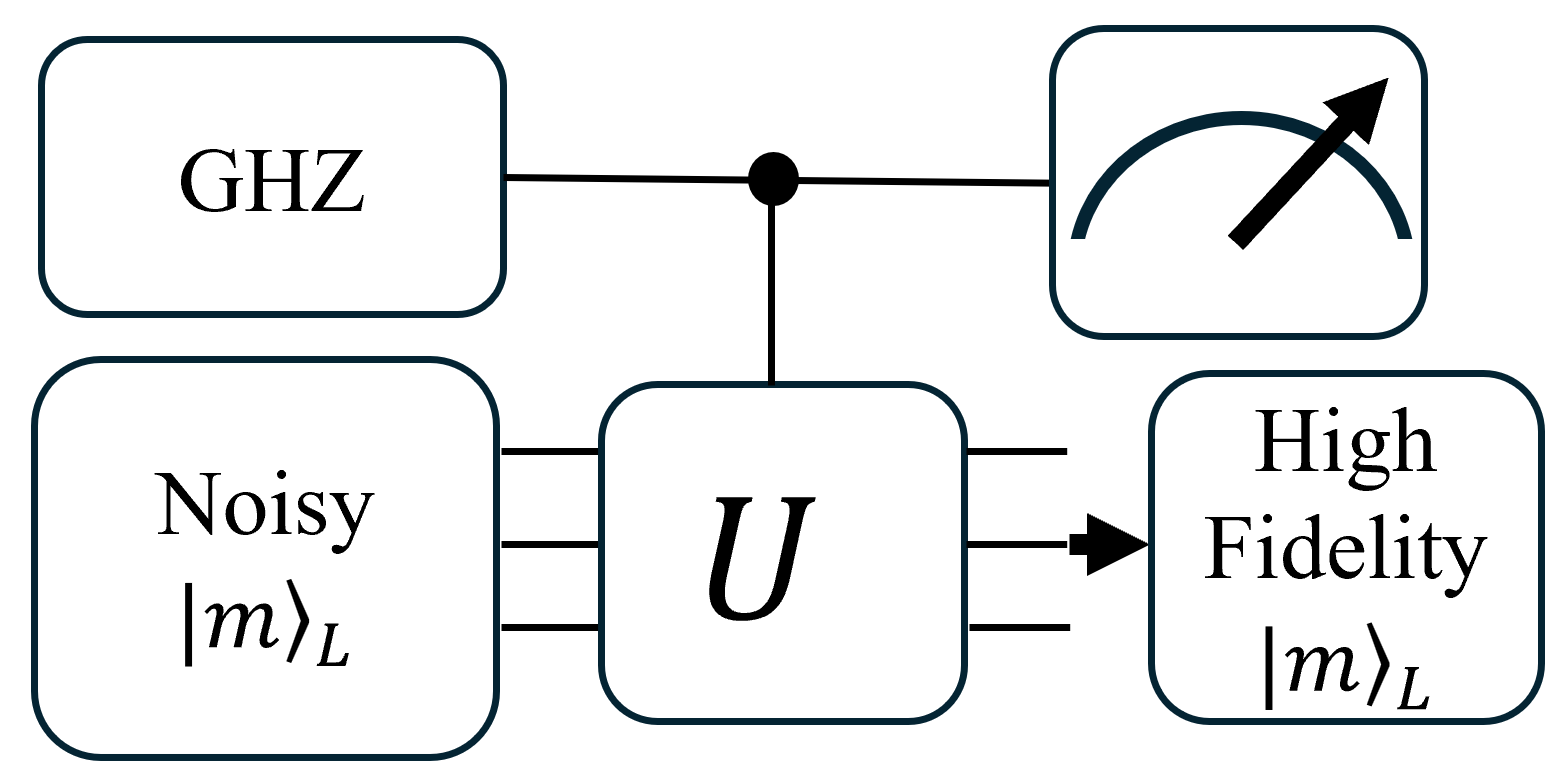}
\caption{\textbf{The phase kickback measurement for arbitrary $\ket{m}_L$.} The $N$ qubits GHZ state $\ket{\psi}=\frac{1}{\sqrt{2}}\left ( \ket{0}^{\bigotimes N}+\ket{1}^{\bigotimes N}  \right )$ are prepared to check the quality of $\ket{m}_L$ by applying tranversally control $U$ gates with target qubits which are the data qubits from the code state and the control qubits which are the qubit of GHZ state. For the final measurement, one can directly measure all ancilla qubits on $X$ basis and check GHZ state's parity. If the final measurement outcome on GHZ state gives an odd parity result, indicating wrong logical state $\ket{m}_L$ preparation, one can discard the result. On the other hand, an even parity result means the high fidelity logical state $\ket{m}_L$ preparation. Alternatively, one can unentangle GHZ state back into all $Z$ product state and measure all qubits in $Z$ basis. If there are any non-zero results, it means the state is corrupted by noise. One can directly abandon the shot.}
\label{fig:dp_check}
\end{figure}

\begin{table}[h]
\centering
\begin{tabular}{l c c}
\toprule
 & $V$ & Final logical error rate \\
\midrule
 Protocol 1 3D LS & 45443 &  $1.5 \times 10^{-5}$ \\
Protocol 2 3D LS & 47181 & $1.5 \times 10^{-5}$ \\
Protocol 1 2D LS & 46170 & $1.5 \times 10^{-5}$ \\
Protocol 2 2D LS & 46733 & $1.5 \times 10^{-5}$ \\
\bottomrule
\end{tabular}
\caption{\textbf{The expected spacetime volume per successful shot for protocol $1$ and $2$} The expected spacetime volume per successful shot to achieve certain logical error rate after the complementary gap based postselection for combination of cultivation and transversal non-Clifford gate with 3D lattice surgery (LS) with idling noise. The more details about the expected spacetime volume are in appendix~\ref{sec: sp_tv}.}
\label{tab:ls_error_rates_transversal}
\end{table}

\section{Conclusions}
In this paper, we adopt a $d=3$ CS protocol for magic-state preparation in the idling noise model and compare with MSC. We observe that magic-state preparation using transversal non-Clifford gates shows strong consistency in performance between $S\ket{+}_L$ and $T\ket{+}_L$ preparations with and without idling noise. Although the state vector simulation result has large uncertainty in logical infidelity, it still shows that the CS protocol is a strong candidate for magic state preparation for near term devices, especially for all-to-all connectivity devices such as trapped-ion and neutral atom devices. Moreover, we provide the corresponding growth methodology to grow from the color code into the rotated surface code. In particular, inspired by Ref.~\cite{Hirano2025}, we establish a protocol for LS between the doubled color code and the rotated surface code to reduce overhead. Additionally, we extend $d=3$ CS based magic state preparation to $d=5$. We show that without idling noise, the infidelity of the $d=5$ CS based preparation for $S\ket{+}_L$ can reach $3 \times 10^{-9}$. 

Finally, we also study the combination of a non-Clifford transversal gate and a double phase kickback check, a component of MSC. There are two ways to combine these two components together: (1) one using a double kickback phase check for $\ket{+}_L$ and then applying the transversal non-Clifford gate to prepare the magic state. And (2) applying the transversal non-Clifford gate first and implementing the double phase kickback check for $T\ket{+}_L$. Compared to the original CS protocol with only transversal $T$ gate, the infidelity of the first protocol with the double kickback phase check is improved by $10\%$. In contrast, the second protocol cannot improve performance relative to the original CS protocol.

This work provides a more comprehensive numerical simulation of magic state preparation using transversal non-Clifford gates, including growth into larger surface code and the $d=5$.  simulation. The result shows that magic state preparation using transversal non-Clifford gates is another near-term device friendly option, especially for neutral atom and trapped-ion devices. Moreover, the corresponding method can be adopted into some magic state preparation methods such as unfolded distillation~\cite{Ruiz2025} and low overhead distillation in the color code~\cite{Lee2025}. Furthermore, an adapter can also be constructed that connects a $2$D color code with a quantum low density parity check (qLDPC) code and surface code to teleport the magic state from the $2$D color code to the qLDPC code~\cite{Xu2025}. 

%Although this work extends CS based magic state preparation to $d=5$, we do not study the $d=5$ case under the idling noise model since parallelizing the $d=5$ color code cycle and surface code cycle is complicated and extending $3$D LS to $d=5$ without losing too many shots after the postselection remains unexplored. 
%In addition, another $d=5$ $3$D tetrahedral color code also has transversal $T$ gate and can be switched into $d=5$ $2$D color code. Those are interesting further explorations that we leave for future work.    

% 

\begin{acknowledgments}
We acknowledge valuable discussions with Chen Zhao, Pei-Kai Tsai, and Kaavya Sahay. This work was supported by the NNSA ASC Beyond Moore's Law project (I.C.C. and A.T.S.). The LANL designation for this manuscript is LA-UR-26-24734. 
The work is funded in part by NSF grants 1818914, 2325080 (with a subcontract to NC State University from Duke University), 2120757 (with a subcontract to NC State University from the University of Maryland) (HPP and HZ). We acknowledge the computing resources provided by North Carolina State University High Performance Computing Services Core Facility (RRID:SCR\_022168).  

\appendix
\section{Magic state cultivation}
MSC's procedure~\cite{Gidney2024} consists of three stages: injection, cultivation, and escape. For the injection stage, one can use any state injection method to prepare a noisy magic state. For the cultivation stage, the core idea is to check the quality of the magic state $\ket{m}_L$ via a phase kickback measurement (the circuit shown in Fig.~\ref{fig:dp_check}). At the beginning of MSC, a noisy magic state and a GHZ is prepared. One can implement a transversal control unitary gate $CU$ on the GHZ state (control qubits) and magic state (target qubits). The corresponding unitary should be a transversal gate for the cultivated code and satisfies
\begin{align}\label{eq:u_check}
    U\ket{m}_L = \left ( +1 \right )\ket{m}_L. 
\end{align}
Hence, the GHZ state remains invariant if no errors occur. On the other hand, if errors occur, the GHZ state can be corrupted to other states. Thus, one directly measures each qubit of the GHZ state in the $X$ basis and checks the corresponding parity. If any odd parity results are given, it indicates that an error has occurred. Alternatively, one can also unentangle the GHZ state into a product state in the $Z$ basis and measure each qubit. If measurements give a non-zero outcome, this also means that an error happened. Hence, one discards the entire shot and repeats the process until success.   

Since magic states, $T\ket{+}_L$, are commonly investigated, $U$ is usually chosen to be $H_{XY}=\frac{X+Y}{2}$ so that one can check $T\ket{+}_L$'s quality. The chosen cultivated code must support a transversal or fold transversal $H_{XY}$ gate. The self-dual quantum CSS code family is a good candidate for $T\ket{+}_L$ cultivation. The fold transversal $H_{XY}$ gate is used on the unrotated surface code and composed of the transversal $X$ gate and fold transversal $S$ gate. With the sequence of operators, $SX$, the magic state becomes 
\begin{align}\label{eq:u_check}
    SXT\ket{+}_L =e^{i\frac{\pi}{4}}T\ket{+}_L,
\end{align}
which subsequently generates a phase on the GHZ state. One can apply $T^\dagger$ to cancel this phase after disentangling the GHZ state. Beyond $H_{XY}$ checks, only the $U=CX$ are investigated for the magic state $\ket{CX}_L\equiv \frac{\ket{0,+}_L+\ket{0,-}_L+\ket{1,+}_L}{\sqrt{3}}$~\cite{Vaknin2025}, which is the resource state for the logical Toffoli implementation~\cite{Dennis2001}. 

After the phase kickback measurement, known as the cultivation stage~\cite{Gidney2024}, the cultivated code must be expanded into the large code so that the high-fidelity state is not significantly affected by memory noise with quantum error correction. For the cultivated $2$D color code, one can expand it to the surface code using grafting or LS, which we discussed in the main text. For the rotated surface code, one can expand it using unitary growth, LS, or Li's protocol~\cite{Li2015}. After the escape stage, one can use a complementary gap calculated from the decoder or other soft postselection methods based on the syndrome measurement outcomes to filter out some logical-error-prone results. 

According to Ref.~\cite{Gidney2024}, $S\ket{+}_L$ and $T\ket{+}_L$ MSC on the $d=3$ $2$D color code give different performance. Moreover, as the noise scale decreases, the difference between the two MSC logical errors becomes larger. Here, to further understand the scaling of $2$D color code MSC for $S\ket{+}_L$ and $T\ket{+}_L$, we test their performance under two different noise models, uniform noise models with and without idling, with varying noise levels. We implement MSC followed by $3$ rounds of noisy syndrome measurement, a round of ideal syndrome measurements, and an ideal inverse logical T gate or S gate. We then estimate the logical infidelity using the revival rate from the result, and also discard any shots with syndrome errors.

\begin{figure}[t]
\centering
\includegraphics[width=1\linewidth]{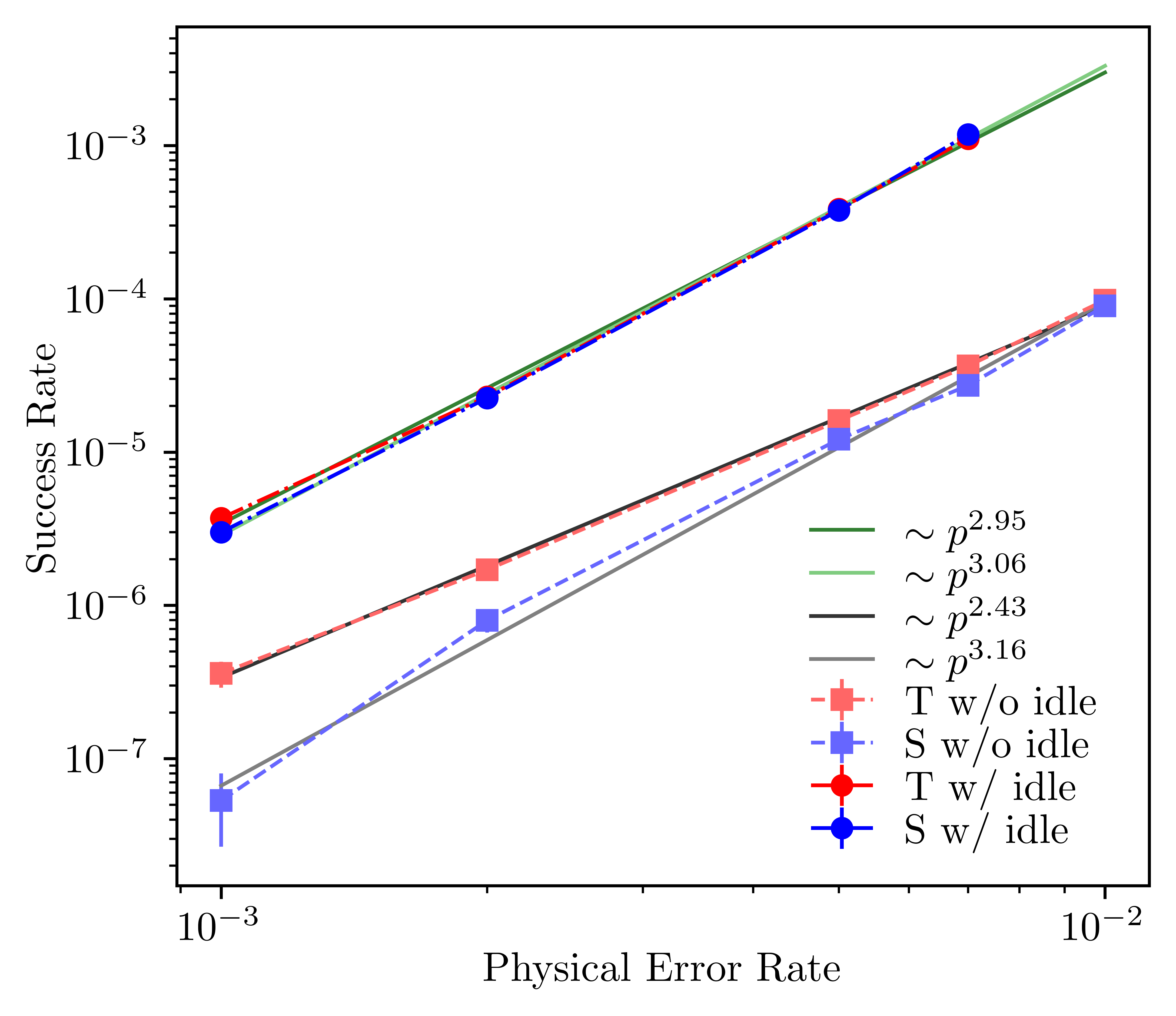}
\caption{\textbf{Logical infidelities of magic state cultivation.} The logical infidelities of T gate and S gate MSC are estimated under a uniform noise model with (w/ idle), without (w/o idle) ideal noise, and with varied noise levels. The exponents from the fitting lines are calculated using $\mathbf{Scipy.state.linregress}$ function. Error bars correspond to the standard error of the mean.} %Result is different from Google's since we just use 14 qubits for MSC while Google adds an extra qubit for GHZ check.
\label{fig:Fig1_msc}
\end{figure}

As Fig.~\ref{fig:Fig1_msc} shows, the logical infidelity discrepancy between $T\ket{+}_L$ and $S\ket{+}_L$ becomes larger as the noise level decays. Especially without idling noise, the gap is more significant. At noise level $p=10^{-3}$, the logical error rate of $T\ket{+}_L$ is roughly $6$ times as that of $S\ket{+}_L$. The result shows that the assumption of a factor of $2$ difference between $T\ket{+}_L$ and $S\ket{+}_L$ doesn't always hold. Moreover, the scalings of  $T\ket{+}_L$ and $S\ket{+}_L$ MSC logical infidelity varying with physical noise level are distinct. Especially, with the uniform noise model without idling noise, the fitting exponent of $S\ket{+}_L$ MSC is $3.16$ while that of $T\ket{+}_L$ MSC is $2.42$.  

Based on this result, it is necessary to have further numerical evidence or analytical calculation for magic cultivation on the $2$D color code under different noise models and with $d=5$ as well. Moreover, it also suggests that other MSCs on the surface code~\cite{Chen2025,Vaknin2025,Sahay2025,Claes2025} may require more numerical simulation or analytical calculation to verify their $T\ket{+}_L$ MSC result, especially for $d=5$ MSC.  Although Ref.~\cite{Li2025} demonstrates ungrown $T\ket{+}_L$ MSC with $d=5$ and shows its logical error rate are roughly as $8$ times as that of $S\ket{+}_L$ MSC, it still requires more samples of simulation to ensure the scaling of $T\ket{+}_L$ MSC and the logical error difference between $S\ket{+}_L$ and $T\ket{+}_L$ MSCs due to the large statistical uncertainty in the numerical result. Moreover, the discrepancy between $T\ket{+}_L$ and $S\ket{+}_L$ MSC behaves differently under different noise models such as for models without idling noise, which are common in ion trap and neutral atom system. 

\begin{table*}[]
\begin{tabular}{|c|c|c|}
 \hline
Process & Spacetime Volume & Description \\ \hline \
$[[15,1,3]]$ $\ket{+}_L$ preparation & $23\times8+15\times2$ & 2 4-qubit GHZ states for checking $\ket{+}_L$ \\ \hline \
$[[7,1,3]]$ $\ket{0}_L$ preparation & $7\times2+9\times6$ & 2 ancilla qubita for checking $\ket{0}_L$ \\ \hline \
Z type Stabilizers measurements & $35\times8$ & 10 Bell states for Z-type syndrome extraction \\ \hline \
X type Stabilizers measurements & $23\times8$ & 4 Bell states for X-type syndrome extraction (only for protocols 1 \& 2)\\ \hline \
$15$q GHZ double phase kickback check & $11\times13$ ($13\times30$) & To check $[[15,1,3]]$ magic state (only for protocols 1 \& 2) \\ \hline \
Transversal T & 15  & Apply $T$ and $T^\dagger$ on each physical qubit of [[15,1,3]] code \\ \hline \
Logical $X$ measurement on $[[15,1,3]]$ & 15  & Measure each physical qubit in $X$ basis \\ \hline \
Transversal CNOT & 22  & Transversal CNOT between Steane code and doubled color code \\ \hline \
$2$D Lattice Surgery ($d_i=3$ to $d_f=11$) & $255\times21+241\times6$   & $3$ rounds of surface code syndrome measurement \\ \hline \
$3$D Lattice Surgery ($d_i=3$ to $d_f=11$) & $277\times24+241\times6$   & $3$ rounds of surface code syndrome measurement\\ \hline \
$10$ rounds of syndrome measurement & $241\times6\times10$   & Waiting for complementary gap calculation \\ \hline
\end{tabular}
\caption{\textbf{The spacetime volume of different parts of magic state preparation 
.} For the double phase kickback check, protocol $1$ dosen't need layers of phase gate to cancel the phase. The corresponding spacetime volume is $11\times30$. In contrast, protocol $2$ needs layers of $T$ and $T^\dagger$ gates to cancel the extra phase. Hence, the total space time volume is $13\times30$. Additionally, transversal CNOT is only used for the $2$D LS. }
 \label{tab: spt}
\end{table*}

\section{Approximate Noisy Circuit}
\label{sec:approx_cir}
In our case, Z stabilizer measurement requires 20 ancillary qubits. With data qubits in the [[15,1,3]] code, the entire circuit simulation for the transversal $T$ gate requires $35$ qubits, which is difficult to simulate using the state vector simulator. Thus, to tackle this problem, we implement each $Z$ stabilizer measurement individually with a Bell pair and locally apply depolarizing noise to the qubits involved in the stabilizer measurements. For the $X$ stabilizers, we implement the same strategy.    

We note that the performance of this approximate circuit differs from that of the original circuit with $35$ physical qubits in total. In particular, in the approximate noisy circuit, the accumulated idle noise from each individual stabilizer measurement induces an amount of logical noise. On the other hand, without idle noise, the logical error is not affected by the circuit depth. As Fig.~\ref{fig:Fig2_transversal} shows, the result of Clifford simulation is consistent with that of state vector simulation for both $T\ket{+}_L$ and $S\ket{+}_L$ preparation.    

\begin{figure}[htb]
\centering
\includegraphics[width=0.7\linewidth]{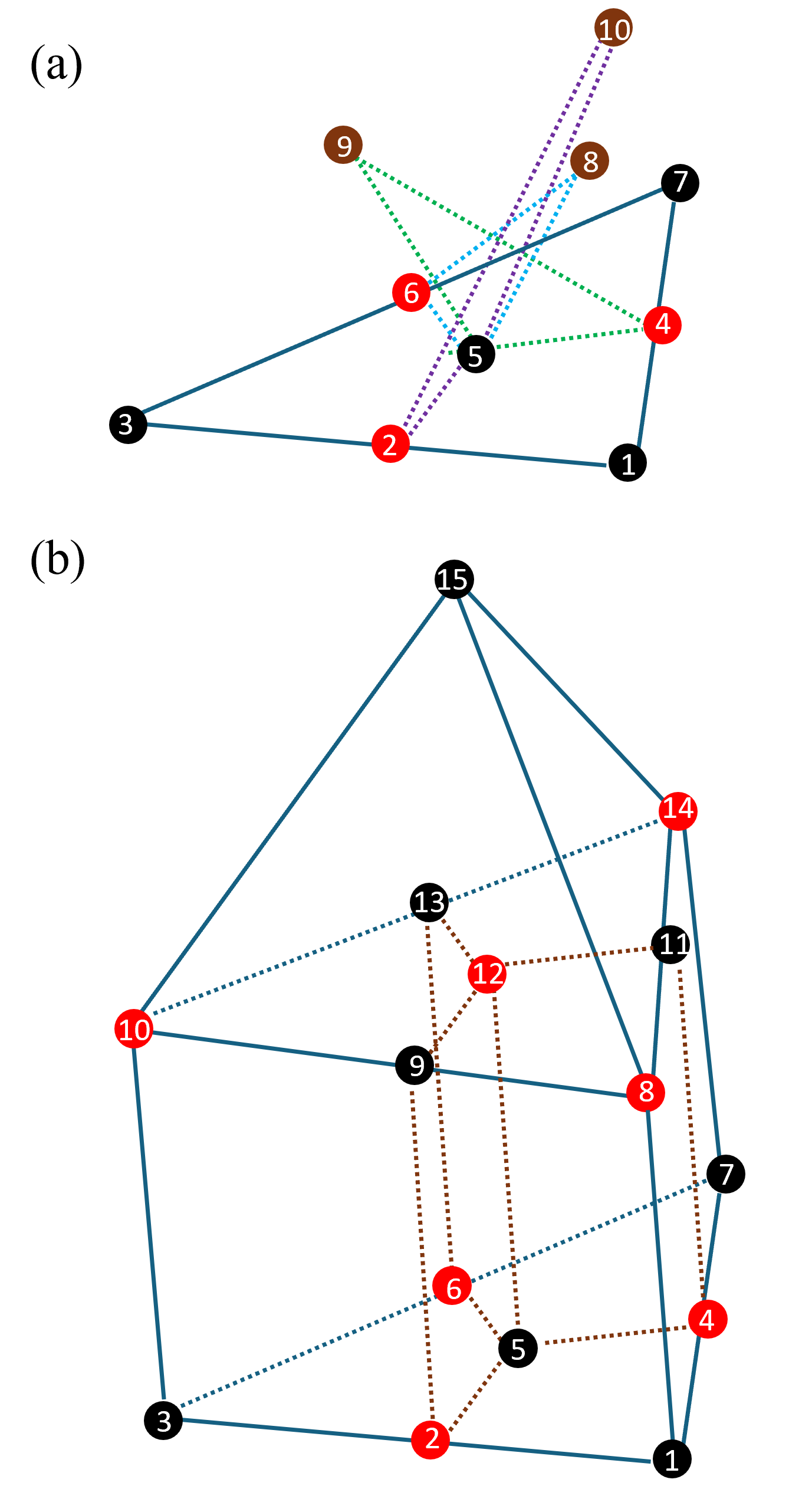}
\caption{\textbf{The visualization of Vasmer-Kubica code and the $3$D [[15,1,3]] color code.} (a) [[10,1,2]] Vasmer-Kubica code. (b) [[15,1,3]] $3$D color code (also known as quantum Reed-Muller code). To implement a transversal $T$ gate on a $3$D color code, one can simultaneously apply $T^\dagger$ gates on all of the black qubits and $T$ gates on all of the red qubits. On the other hand, for the transversal $T$ gate on the Vasmer-Kubica code, one also applies $T^\dagger$ gates on all of the black qubits, $T$ gates on all of the red qubits, and a $CCZ$ gate on the brown qubits.}
\label{fig:15_1_3}
\end{figure}

\section{Space time volume evaluation}
\label{sec: sp_tv}
Here, we provide more details on spacetime volume estimation for the magic state preparation methods that we discuss above. Each protocol above can be decomposed into different components. We optimize the quantum circuits for the magic state preparation and estimate the corresponding individual component's spacetime volume.

In Table~\ref{tab: spt}, detailed spacetime volume estimates for magic state preparation are shown. For the initialization of the $\ket{+}_L$ on $[[15,1,3]]$ code, without the flag qubit's check, the circuit depth is $5$. Since we use $2$ 4-qubit GHZ states to check the logical state, we can partially parallelize the GHZ state preparation and the $[[15,1,3]]$ $\ket{+}_L$ preparation. Similarly, we also use a Bell pair to check the Steane code's preparation. Thus, we also partially parallelize preparation of Bell pair and $\ket{0}_L$ on $[[7,1,3]]$ code. To optimize the parallelization, we also use Bell pairs to extract the syndrome for $Z$ and $X$ stabilizers. Only protocols $1$ and $2$ involve the GHZ double phase kickback check and $X$ stabilizer measurement. Therefore, the total spacetime volumes of protocols $1$ and $2$ are larger than those without the double phase kickback check. During LS, we implement the $Z$ or $X$ color code stabilizer measurements to parallelize the surface code syndrome measurement cycle. Since $X$ or $Z$ syndrome measurements in $2$D (doubled) color code require one (two, resp.) more circuit depth than surface code cycle, the total circuit depth for each round of LS is $7$ ($8$) instead of $6$. After the final measurement of the color code, a single round of surface code stabilizer measurement is applied to stabilizers at the boundary of the surface code. In total, for $d=3$, there are $3$ rounds of LS. However, for the color code, there are only two rounds of stabilizer measurement (including the final measurement of all qubits of the color code in $X$ basis). After LS, we also implement an extra $10$ rounds of syndrome measurement on the surface code in order to wait for the calculation from the decoder for complementary gap based postselection. The main contribution of the spacetime volume comes from these $10$ extra rounds of stabilizer measurements. 

To calculate the expected spacetime volume per successful shot ($V$ in eq.~\ref{eq:stv}), we also obtain the success rate for different parts of the circuit from the protocols we mentioned. Although protocols $1$ and $2$ give different success rates at the end, as Table~\ref{tab:ls_error_rates_transversal} shows, they provide similar $V$ to achieve an infidelity $1.5\times10^{-5}$ for the magic state. 

\section{Doubled color code and Vasmer-Kubica code}
\begin{figure}[htb]
\centering
\includegraphics[width=1.0\linewidth]{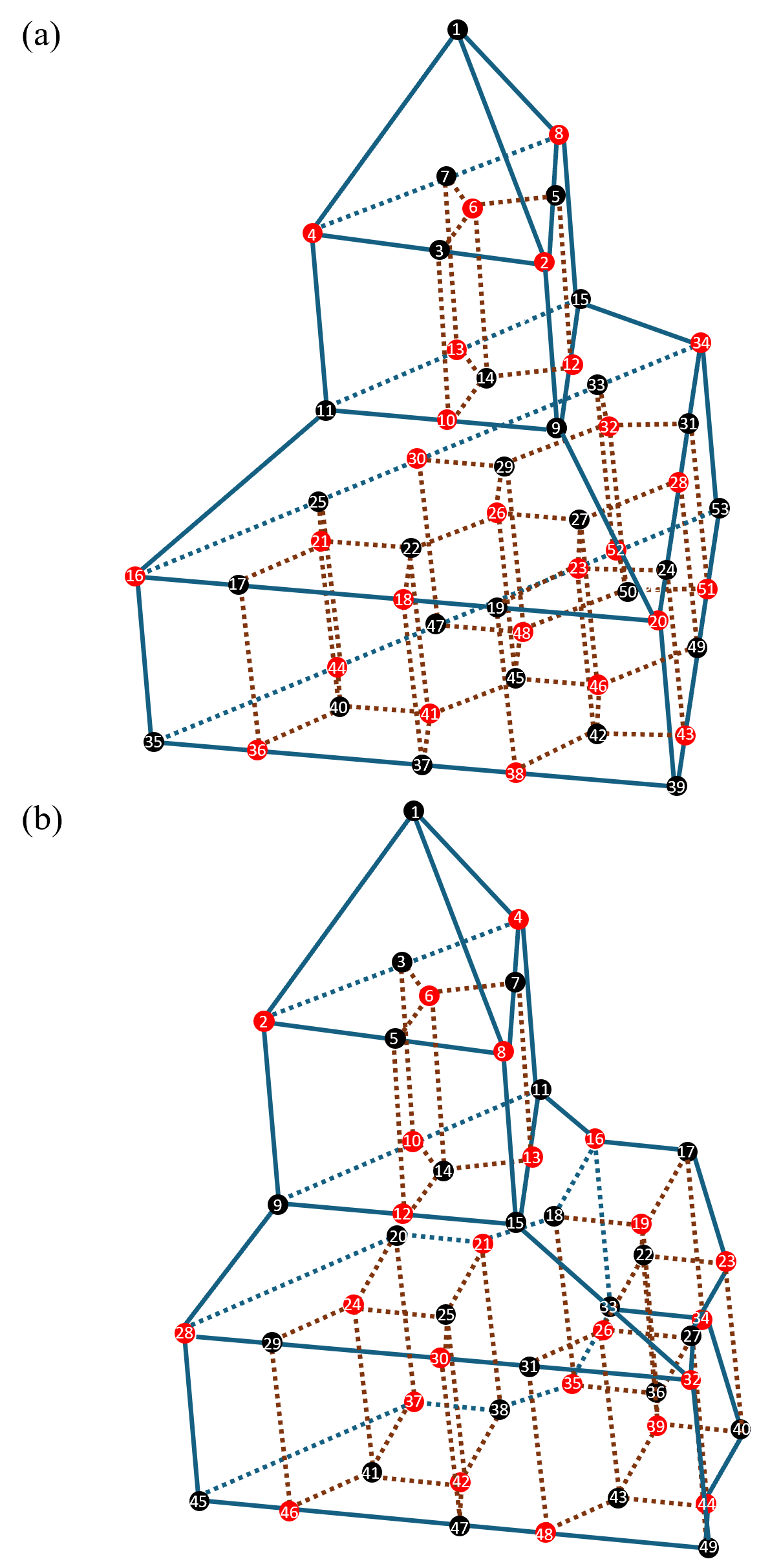}
\caption{\textbf{The visualization of $3$D recursive capped color code.} (a) [[49,1,5]] code constructed from [[17,1,5]] code. (b) [[53,1,5]] code constructed from [[19,1,5]] code. To implement the transversal $T$ gate on those two code, one can apply physical $T$ gates on the black qubits and $T^\dagger$ on the red qubits. The $X$ ($Z$) type of stabilizers are on the cells (faces).}
\label{fig:d5_3S_code}
\end{figure}

The $3$D color code [[15,1,3]] is also known as the Quantum Reed-Muller (QRM) code. It can be obtained by the doubling the construction from the [[7,1,3]] Steane code with fixed gauges. It has $4$ $8$-weight $X$ stabilizers and $10$ $4$-weight $Z$ stabilizers, which are on the cells and the faces, respectively, as seen in Fig.~\ref{fig:15_1_3}(b). The corresponding $X$ checks are
\begin{align}\label{eq:X_s_15}
 S_{X,[[15,1,3]]}=\begin{pmatrix}
 X_1X_2X_4X_5X_8X_9X_{11}X_{12}\\
 X_2X_3X_5X_6X_9X_{10}X_{12}X_{13}\\
 X_4X_5X_6X_7X_{11}X_{12}X_{13}X_{14}\\
 X_8X_9X_{10}X_{11}X_{12}X_{13}X_{14}X_{15}
 \end{pmatrix}
\end{align}
and the $Z$ checks are
\begin{align}\label{eq:Z_s_15}
 S_{Z,[[15,1,3]]}=\begin{pmatrix}
 Z_1Z_2Z_4Z_5\\
 Z_2Z_3Z_5Z_6\\
 Z_4Z_5Z_6Z_7\\
 Z_8Z_9Z_{11}Z_{12}\\
  Z_9Z_{10}Z_{12}Z_{13}\\
  Z_{11}Z_{12}Z_{13}Z_{14}\\
  Z_{10}Z_{13}Z_{14}Z_{15}\\
  Z_{3}Z_{6}Z_{10}Z_{13}\\
  Z_{2}Z_{3}Z_{9}Z_{10}\\
  Z_{1}Z_{2}Z_{8}Z_{9}
 \end{pmatrix}
\end{align}
The $Z$ logical operators exist on the edge of the tetrahedron. Hence, the corresponding code distance for $Z$ errors is $d_z=3$. On the other hand, the $X$ stabilizers are located at the face of the tetrahedron. The code distance for $X$ errors is $d_x=7$.    

 \begin{table*}[]
\begin{tabular}{|c|c|c|}
 \hline
             & Z & X \\ \hline
$[[49,1,5]]$ & $\begin{pmatrix}
 Z_2Z_3Z_5Z_6\\
 Z_3Z_4Z_6Z_7\\
 Z_5Z_6Z_7Z_8\\
 Z_9Z_{10}Z_{12}Z_{14}\\
  Z_{10}Z_{11}Z_{13}Z_{14}\\
  Z_{11}Z_{12}Z_{13}Z_{14}\\
  Z_1Z_2Z_5Z_8\\
  Z_2Z_5Z_9Z_{12}\\
  Z_2Z_3Z_9Z_{10}\\
  Z_3Z_4Z_{10}Z_{11}\\
  Z_{16}Z_{17}Z_{18}Z_{19}\\
  Z_{33}Z_{34}Z_{35}Z_{36}\\
  Z_{17}Z_{19}Z_{22}Z_{23}\\
  Z_{34}Z_{36}Z_{39}Z_{40}\\
  Z_{22}Z_{23}Z_{26}Z_{27}\\
  Z_{39}Z_{40}Z_{43}Z_{44}\\
  Z_{26}Z_{27}Z_{31}Z_{32}\\
  Z_{43}Z_{44}Z_{48}Z_{49}\\
  Z_{18}Z_{19}Z_{22}Z_{26}Z_{31}Z_{30}Z_{25}Z_{21}\\
  Z_{35}Z_{36}Z_{39}Z_{43}Z_{48}Z_{47}Z_{42}Z_{38}\\
  Z_{20}Z_{21}Z_{24}Z_{25}\\
  Z_{37}Z_{38}Z_{41}Z_{42}\\
  Z_{24}Z_{25}Z_{29}Z_{30}\\
  Z_{41}Z_{42}Z_{46}Z_{47}\\
  Z_{20}Z_{24}Z_{28}Z_{29}\\
  Z_{37}Z_{41}Z_{45}Z_{46}\\
  Z_{9}Z_{12}Z_{15}Z_{28}Z_{29}Z_{30}Z_{31}Z_{32}\\
  Z_{20}Z_{21}Z_{37}Z_{38}\\
  Z_{20}Z_{28}Z_{37}Z_{45}\\
  Z_{18}Z_{19}Z_{35}Z_{36}\\
  Z_{19}Z_{22}Z_{36}Z_{39}\\
  Z_{22}Z_{26}Z_{39}Z_{43}\\
  Z_{26}Z_{31}Z_{43}Z_{48}\\
  Z_{31}Z_{32}Z_{48}Z_{49}\\
  Z_{28}Z_{29}Z_{45}Z_{46}
 \end{pmatrix}$
&  $\begin{pmatrix}
 X_1X_2X_3X_4X_5X_6X_7X_8\\
 X_2X_3X_5X_6X_9X_{10}X_{12}X_{14}\\
 X_3X_4X_6X_7X_{10}X_{11}X_{14}X_{13}\\
 X_5X_6X_7X_8X_{11}X_{14}X_{13}X_{15}\\
 S_{24}\\
 X_{16}X_{17}X_{18}X_{19}X_{33}X_{34}X_{35}X_{36}\\
 X_{17}X_{19}X_{22}X_{23}X_{34}X_{36}X_{39}X_{40}\\
 X_{22}X_{23}X_{26}X_{27}X_{39}X_{40}X_{43}X_{44}\\
 X_{26}X_{27}X_{31}X_{32}X_{43}X_{44}X_{48}X_{49}\\
 X_{18}X_{19}X_{21}X_{26}X_{31}X_{30}X_{25}X_{21}X_{35}X_{36}X_{39}X_{43}X_{48}X_{47}X_{42}X_{38}\\
 X_{20}X_{21}X_{24}X_{25}X_{37}X_{38}X_{41}X_{42}\\
 X_{24}X_{25}X_{29}X_{30}X_{41}X_{42}X_{46}X_{47}\\
 X_{20}X_{24}X_{28}X_{29}X_{37}X_{41}X_{45}X_{46}
 \end{pmatrix}$ \\ \hline
\end{tabular}
\caption{\textbf{The stabilizers of $[[49,1,5]]$. $S_{24}=X_9X_{10}X_{11}X_{12}X_{14}X_{13}X_{15}X_{16}X_{17}X_{18}X_{19}
X_{20}X_{21}X_{22}X_{23}X_{24}X_{25}X_{26}X_{27}X_{28}\\X_{29}X_{30}X_{31}X_{32}$} The labels correspond to the label in Fig.~\ref{fig:d5_3S_code} (b).}
\end{table*}

The Vasmer-Kubica code is a [[10,1,2]] code morphed from the [[15,1,3]] quantum Reed-Muller code with a subset of qubits that defines [[8,3,2]] code~\cite{Vasmer2022}. It has $3$ $3$-weight $Z$ stabilizers and $3$ $4$-weight $Z$ stabilizers. As Fig.~\ref{fig:15_1_3}(a) shows, The corresponding $Z$ checks are given by 
\begin{align}\label{eq:Z_s_10}
 S_{Z,[[10,1,2]]}=\begin{pmatrix}
 Z_1Z_2Z_4Z_5\\
 Z_2Z_3Z_5Z_6\\
 Z_4Z_5Z_6Z_7\\
 Z_2Z_5Z_{10}\\
 Z_4Z_5Z_9\\
Z_5Z_6Z_8\end{pmatrix} \; .
\end{align}
\end{acknowledgments}
There are $3$ $5$-weight $X$ stabilizers for the Vasmer-Kubica code 
\begin{align}\label{eq:X_s_10}
 S_{X,[[10,1,2]]}=\begin{pmatrix}
 X_1X_2X_4X_5X_8\\
 X_2X_3X_5X_6X_9\\
 X_4X_5X_6X_7X_{10}
 \end{pmatrix} \; .
\end{align}
The code also has $Z$ logical operators $Z_1Z_8$, $Z_3Z_9$, $Z_7Z_{10}$. Hence, its $Z$ code distance is $2$. On the the other hand, there is the $X$ logical operator $X_1X_2X_3X_4X_5X_6X_7$. The code distance for $X$ error is $7$. In Fig.~\ref{fig:15_1_3}(a), we see that the code supports transversal $T$ by applying physical $T$ gates on the black qubits, $T^\dagger$ gates on the red qubits, and a $CCZ$ gate on the brown qubits.

 \begin{table*}[]
\begin{tabular}{|c|c|c|}
 \hline
             & Z & X \\ \hline
$[[53,1,5]]$ & $\begin{pmatrix}
 Z_2Z_3Z_5Z_6\\
 Z_3Z_4Z_6Z_7\\
 Z_5Z_6Z_7Z_8\\
 Z_9Z_{10}Z_{12}Z_{14}\\
  Z_{10}Z_{11}Z_{13}Z_{14}\\
  Z_{11}Z_{12}Z_{13}Z_{14}\\
  Z_1Z_2Z_5Z_8\\
  Z_2Z_5Z_9Z_{12}\\
  Z_2Z_3Z_9Z_{10}\\
  Z_3Z_4Z_{10}Z_{11}\\
  Z_{16}Z_{17}Z_{25}Z_{21}\\
  Z_{35}Z_{36}Z_{44}Z_{40}\\
  Z_{17}Z_{18}Z_{21}Z_{22}\\
  Z_{36}Z_{37}Z_{40}Z_{41}\\
  Z_{18}Z_{19}Z_{23}Z_{27}Z_{26}Z_{22}\\
  Z_{37}Z_{38}Z_{42}Z_{46}Z_{45}Z_{41}\\
  Z_{19}Z_{20}Z_{23}Z_{24}\\
  Z_{38}Z_{39}Z_{42}Z_{43}\\
  Z_{21}Z_{22}Z_{26}Z_{29}Z_{28}Z_{25}\\
  Z_{40}Z_{41}Z_{45}Z_{48}Z_{47}Z_{44}\\
  Z_{23}Z_{24}Z_{30}Z_{27}\\
  Z_{42}Z_{43}Z_{49}Z_{46}\\
  Z_{31}Z_{32}Z_{30}Z_{27}Z_{26}Z_{29}\\
  Z_{50}Z_{51}Z_{49}Z_{46}Z_{45}Z_{48}\\
  Z_{22}Z_{26}Z_{30}Z_{31}\\
  Z_{32}Z_{31}Z_{28}Z_{29}\\
  Z_{34}Z_{32}Z_{31}Z_{33}\\
  Z_{53}Z_{52}Z_{50}Z_{51}\\
  Z_{9}Z_{10}Z_{11}Z_{16}Z_{17}Z_{18}Z_{19}Z_{20}\\
  Z_{19}Z_{20}Z_{38}Z_{39}\\
  Z_{16}Z_{17}Z_{35}Z_{36}\\
  Z_{16}Z_{25}Z_{35}Z_{44}\\
  Z_{21}Z_{22}Z_{40}Z_{41}\\
  Z_{22}Z_{36}Z_{41}Z_{45}\\
  Z_{26}Z_{27}Z_{45}Z_{46}\\
  Z_{29}Z_{26}Z_{45}Z_{48}\\
  Z_{29}Z_{31}Z_{48}Z_{50}\\
  Z_{23}Z_{27}Z_{42}Z_{46}
 \end{pmatrix}$
&  $\begin{pmatrix}
 X_1X_2X_3X_4X_5X_6X_7X_8\\
 X_2X_3X_5X_6X_9X_{10}X_{12}X_{14}\\
 X_3X_4X_6X_7X_{10}X_{11}X_{14}X_{13}\\
 X_5X_6X_7X_8X_{11}X_{14}X_{13}X_{15}\\
 S_{26}\\
 X_{16}X_{17}X_{25}X_{21}X_{35}X_{36}X_{44}X_{40}\\
 X_{17}X_{18}X_{21}X_{22}X_{36}X_{37}X_{40}X_{41}\\
 X_{18}X_{19}X_{23}X_{27}X_{26}X_{22}X_{37}X_{38}X_{42}X_{46}X_{45}X_{41}\\
 X_{19}X_{20}X_{23}X_{24}X_{38}X_{39}X_{42}X_{43}\\
 X_{21}X_{22}X_{26}X_{29}X_{28}X_{25}X_{40}X_{41}X_{45}X_{48}X_{47}X_{44}\\
 X_{23}X_{24}X_{30}X_{27}X_{42}X_{43}X_{49}X_{46}\\
 X_{31}X_{32}X_{30}X_{27}X_{26}X_{29}X_{50}X_{51}X_{49}X_{46}X_{45}X_{48}\\
 X_{52}X_{50}X_{47}X_{48}X_{33}X_{31}X_{28}X_{29}\\
 X_{34}X_{32}X_{31}X_{33}X_{53}X_{52}X_{50}X_{51}
 \end{pmatrix}$ \\ \hline
\end{tabular}
\caption{\textbf{The stabilizers of $[[53,1,5]]$.} $S_{26}=X_9X_{10}X_{11}X_{12}X_{14}X_{13}X_{15}X_{16}X_{17}X_{18}X_{19}
X_{20}X_{21}X_{22}X_{23}X_{24}X_{25}X_{26}X_{27}X_{28}\\X_{29}X_{30}X_{31}X_{32}X_{33}X_{34}$ is a $26$-weight stabilizer. The labels correspond to the label in Fig.~\ref{fig:d5_3S_code} (a).} 
\end{table*}

When $d=5$, one can construct the $d=5$ doubled code using double construction. Based on the double construction protocol, one needs a $d=3$ tri-orthogonal code $[[n_{tri},1,3]]$ and $d=5$ quantum with code parameters $[[n_{sd},1,5]]$ to construct the doubled code $[[n_{tri}+2n_{sd},1,5]]$. For the tri-orthogonal code, we select the QRM code $[[15,1,3]]$ described above. On the other hand, the $d=5$ $2$D color code $[[19,1,5]]$ and the $[[17,1,5]]$ code can be the self-dual code to construct the doubled color code $[[53,1,5]]$ and $[[49,1,5]]$ (shown in Fig.~\ref{fig:d5_3S_code}) respectively. $[[49,1,5]]$ has $35$ $Z$ stabilizers, which are located on the faces and given in the table, and $X$ stabilizers, which are on the cells. Since these codes still retain their $2$D color code $Z$ stabilizers, the codes still support the CS to $d=5$ $2$D color code using a one way transversal CNOT.       

\clearpage
\bibliography{refs}
\end{document}